\documentclass[12pt]{iopart}

\usepackage{hyperref}
\usepackage{footmisc}

\usepackage[utf8]{inputenc}
\usepackage{amssymb}
\expandafter\let\csname equation*\endcsname\relax
\expandafter\let\csname endequation*\endcsname\relax

\usepackage{bbold}
\usepackage{amsmath}

\usepackage{slashed}
\usepackage{nicefrac}

\usepackage{braket}

\usepackage{hyperref}
\usepackage[nosort]{cite}

\usepackage{iopams}  
\usepackage{hyperref}
\usepackage{cite}

\expandafter\let\csname equation*\endcsname\relax
\expandafter\let\csname endequation*\endcsname\relax

\usepackage{multicol,color,textcomp,tikz,url}

\usepackage{xy}
\xyoption{all}

\newcommand{\HA}{{\rm H}} 
 
\newcommand{\fs}{\footnotesize} 
 
\newcommand{\Li}{{\rm Li}}

\newcommand*\pFqskip{8mu}
\catcode`,\active
\newcommand*\pFq{\begingroup
	\catcode`\,\active
	\def ,{\mskip\pFqskip\relax}%
	\dopFq
}
\catcode`\,12
\def\dopFq#1#2#3#4#5{%
	{}_{#1}F_{#2}\biggl[\genfrac..{0pt}{}{#3}{#4};#5\biggr]%
	\endgroup
}

\newcommand{\Dx}{D_x}
\newcommand{\Mvec}{{\rm\bf M}}

\newcommand{\SigmaP}{\texttt{Sigma}}
\newcommand{\ep}{\varepsilon}
\newcommand{\vect}[1]{\mathbf{#1}}

\providecommand{\set}{}
\renewcommand{\set}[1]{{\mathbb #1}}

\providecommand{\KK}{}
\renewcommand{\KK}{\mathbb{K}}

\providecommand{\ZZ}{}
\renewcommand{\ZZ}{\mathbb{Z}}

\providecommand{\QQ}{}
\renewcommand{\QQ}{\mathbb{Q}}

\allowdisplaybreaks[4]

\begin{document}
\bibliographystyle{iopart-num}
\newcommand{\eprint}[2][]{\href{https://arxiv.org/abs/#2}{\tt{#2}}}

\begin{flushleft}
		SAGEX-22-05 \hfill {\tt arXiv:2203.13015 [hep-th]}
		\\
        DESY 22--032\\     
        DO--TH 22/07\\
        RISC Report number 22--03
\end{flushleft}

\title[Multi-loop Feynman integrals]{The SAGEX Review on Scattering Amplitudes \\ 
	Chapter 4: Multi-loop Feynman Integrals}

\author{Johannes Bl\"umlein}
\address{Deutsches Elektronen--Synchrotron DESY, Platanenallee 6, 
	15738 Zeuthen, Germany}
\ead{Johannes.Bluemlein@desy.de}
\author{Carsten Schneider}
\address{Johannes Kepler University Linz, Research Institute for Symbolic Computation (RISC), Altenberger Stra\ss{}e 69, A-4040 Linz, 
	Austria}
\ead{Carsten.Schneider@risc.jku.at}

\vspace{10pt}
\begin{indented}
\item[] March, 2022
\end{indented}

\begin{abstract}
The analytic integration and simplification of multi-loop Feynman integrals to special functions 
and constants plays an important role to perform higher order perturbative calculations in the 
Standard Model of elementary particles. In this survey article the most recent and relevant computer 
algebra and special function algorithms are presented that are currently used or that may play an 
important role to perform such challenging precision calculations in the future. They are discussed 
in the context of analytic zero, single and double scale calculations in the Quantum Field Theories of 
the Standard Model and effective field theories, also with classical applications. These calculations 
play a central role in the analysis of precision measurements at present and future colliders to 
obtain ultimate information for fundamental physics.
\end{abstract}

%
%
%
%
%

\section{Introduction}\label{Sec:Introduction}

\vspace*{1mm}
\noindent
The present and upcoming high luminosity results at the Large Hadron Collider (LHC) at CERN, with input of the results 
measured at the $ep$-collider HERA at DESY, yield a big amount of precision data which require 
further fundamental precision 
calculations in perturbative Quantum Chromodynamics (QCD). This also applies to projects in the Future like the EIC 
\cite{Boer:2011fh},
the LHeC \cite{LHeC:2020van,LHeCStudyGroup:2012zhm}, the ILC 
\cite{ECFADESYLCPhysicsWorkingGroup:1997vxr,
ECFADESYLCPhysicsWorkingGroup:2001igx,ILC:2007bjz,ILC:2019gyn}
or CLIC \cite{vanderMeer:1988yu,CLICPhysicsWorkingGroup:2004qvu,Roloff:2018dqu}, the FCC\_ee 
\cite{FCC:2018evy}, and the proton version of the FCC \cite{FCC:2018evy}, also for Quantum 
Electrodynamics (QED).
Many of these outstanding problems can be formulated by large expressions in terms of hundred thousands (and even millions) 
of sophisticated Feynman integrals at higher loop order of presently up to three and four loops, with 
no, a single and two scales, or even
multiscale problems.\footnote{For zero scale problems results are available also at the five loop 
level.} 
Up to one scale, and in much fewer cases at two scales, technologies have been designed to calculate
these integrals analytically over sets of basic functions, the properties of which have been studied to a certain extent.
Furthermore, numerical representations of these building blocks have been derived.

The central objects are $s$-fold multiple integrals of the form
\begin{equation}
\label{Equ:GeneralForm}
F(n,\ep)=\int_0^1\dots\int_0^1f(n,\ep,x_1,\dots,x_s)dx_1\dots\,dx_s
\end{equation}
where the discrete parameter $n$ stands for the Mellin variable, and $\ep = D - 4, \ep \in 
\mathbb{R}, |\ep| \ll 1$, is the dimensional parameter. 
A crucial property is that the integrand $f$ is hyperexponential\footnote{$h(x)$ is hyperexponential (or hypergeometric) in $x$ if $\frac{h'(x)}{h(x)}$ (or $\frac{h(x+1)}{h(x)}$) is a rational function in $x$.} in each of the integration variables $x_i$ $(1\leq i\leq s)$ and hypergeometric in the discrete parameter $n$. In particular, one is interested in calculating the first coefficients of their Laurent series expansion w.r.t.\ $\ep$:
\begin{equation}\label{Equ:EpExpansion}
F(n,\ep)=F_l(n)\ep^{l}+F_{l+1}(n)\ep^{l+1}+\dots+F_r(n)\ep^{r}+O(\ep^{r+1}).
\end{equation}
\noindent 
For $m$-loop Feynman integrals without infrared divergences such expansions start usually at $l=-m$.
In other cases, one obtains $l=-2m$.
Alternatively, one looks for such an $\ep$-expansions for the inverse Mellin transform $f(x,\ep)$ with
\begin{equation}\label{Equ:OtherRepM}
\Mvec[f(x,\ep)](n) = \
F(n,\ep)=\int_{0}^1x^{n-1}f(x,\ep) dx
\end{equation} 
or its power series representation
\begin{equation}\label{Equ:OtherRepPS}
\bar{f}(x,\ep)=\sum_{n=0}^{\infty}F(n,\ep)x^n.
\end{equation} 
During the last decades more and more significant methods have been derived to simplify such Feynman 
integrals. Based on the representation (for instance~\eqref{Equ:EpExpansion}--\eqref{Equ:OtherRepPS}), 
we will present
important tools that are currently used to perform such challenging calculations and discuss the 
associated special function spaces. We further relate these aspects to different precision 
calculations.

It is needless to say that we had to leave out the description of a series of techniques, which are
also important. This concerns a series of aspects, which have been surveyed in Ref.~\cite{Abreu:2022mfk}, 
appearing in the same volume, and has been agreed between the different authors. It concerns e.g.
the use of the symbol \cite{Duhr:2011zq} and specific Hopf algebra structures 
\cite{Connes:1998qv,Connes:1999yr,Connes:2000fe}, which are omnipresent in quantum field theoretic 
calculations. Related to this, many methods found in algebraic and arithmetic geometry are 
applicable \cite{MANIN,Bloch:2005bh,Bloch:2010gk,BROWN11,Golden:2013xva,
Broadhurst:2014fga,Brown:2015fyf,Bonisch:2021yfw}. For mostly numerical methods in use in multi--leg 
calculations we refer to \cite{Heinrich:2020ybq}.

One way to classify the emergence of new 
mathematical structures in quantum field theories is given by the 
study of their differential equations. This is first of all a 
practical issue at the respective loop--level, where these structures 
are recognized and tied up with the respective graph topologies. 
There is in general no all order statement possible ab initio. 
However, as has been found out recently, Calabi--Yau motives play an 
important role here, cf. e.g.~\cite{Bonisch:2021yfw}. A selection 
criterion on what we are focusing on in the following are key 
technologies for multi--loop calculations in the massive case, 
presently to three--loop order in the zero-, single-, and two-scale 
cases. These technologies do synonymously apply to the corresponding 
massless calculations. There, clearly simplifications can be 
obtained using even other technologies, unlike the case in the 
massive case.

In the following we discuss the following key research topics:
\begin{itemize}
	\item \textit{guessing methods} (see Section~\ref{Sec:Guessing}),

	\item \textit{solving linear recurrences and differential equations} (see Section~\ref{Sec:LinearSolver}),
	
	\item \textit{solving coupled systems} (see Section~\ref{Sec:SolveSystems}),
	
	\item \textit{transformation to special integral and sum representations} (see Section~\ref{Sec:Transformation}),

	\item \textit{symbolic summation} (see Section~\ref{Sec:Summation}),
	
	\item \textit{symbolic integration} (see Section~\ref{Sec:Integration}),
		
	\item \textit{the large moment method} (see Section~\ref{Sec:largemoment}),
	
	\item \textit{special function tools} (see Section~\ref{Sec:SpecialFunctions})

        \item and \textit{concrete calculations in the Quantum Field Theories of the Standard Model 
and within effective field theories} (see Section~\ref{sec:CQFT}).
\end{itemize}
\noindent We emphasize that each of the different techniques cannot be considered as a stand-alone 
toolbox. Contrary, they all have to be applied in non-trivial interactions. In particular, based on a 
concrete problem, one has to choose the best tactic among the conglomeration of tools. For further and supplementary aspects we refer also to~\cite{Blumlein:2018cms,
Blumlein:2021pgo,Weinzierl:22} and Chapter~3~\cite{Abreu:2022mfk} of
the SAGEX 
review. We will conclude this survey on multi-loop tools and multi-loop 
calculations in Section~\ref{Sec:Conclusion}.

\section{Guessing methods}
\label{Sec:Guessing}

\vspace*{1mm}
\noindent
Often physical quantities can be evaluated up to a certain precision and one seeks for a mathematical representation that allows one to represent the data in a more compact fashion, to gain further insight and to support further calculations that depend on these quantities. Here we will emphasize two crucial tactics: (1) to predict from a given floating point number (that approximates a real number to very high precision) an alternative representation in terms of special constants and (2) to guess from a finite set of evaluations at integer points a linear recurrence or linear differential equation that satisfies all evaluations at integer points of the physical quantity.

\subsection{Guessing integer relations}
\label{Sec:IntegerRel}

\vspace*{1mm}
\noindent
Using the LLL algorithm~\cite{Lenstra82}, or the PSLQ algorithm introduced in~\cite{PSLQ} and 
substantially improved in~\cite{Ferguson91apolynomial,PSLQImproved}, one can try to solve the 
following problem:
Given a finite set of finite floating point numbers $a_1,\dots,a_n$ with high precision (say $l$ fractional digits), find integers  $z_1,\dots,z_n\in\set Z$ as small as possible in its 
absolute value\footnote{Since any finite floating point number can be written as a rational number, this problem can be always solved if the integers $z_i$ can be arbitrarily large. Thus a solution to the problem might indicate a proper relation among the approximated real numbers if $l$ is large but the values $z_i$ are small.} such that
$$z_1\,a_1+\dots+z_n\,a_n<10^{-m},$$
where $m$ is large. 

In order to obtain further confidence one may apply this method with further precision (i.e., more digits $l$ of the input) and checks if the obtained result remains the same but $m$ gets larger. For instance, suppose that we are given the finite floating point number 
$$a_1 = 5.68700407989058207630312605688168433418849655155997$$
with $l=50$ fractional digits that approximates a real number $r_1\in\set R$. Then one can use, e.g., 
the Mathematica implementation of the PSLQ algorithm to search for 
an alternative representation in terms of $\zeta(2)$, $\zeta(3)$ and $\zeta(5)$. Namely, by activating 
\texttt{FindIntegerNullVector[\{$a_1$, N[Zeta[2], 50], N[Zeta[3], 50], N[Zeta[5], 50]\}]} one obtains 
the result $(z_1,z_2,z_3,z_4)=(-2, 2, 5, 2)$ with $m=49$. Thus we may conjecture that 
$$r_1=\zeta(2)+\frac52\zeta(3)+\zeta(5)$$
holds. We remark that the PSLQ method finds this relations already with 
$l=5$ fractional digits with precision $m=4$. 
Of course, if one is given even more digits of $r_1$, one may check 
if even more digits $m$ agree. Similarly, one may apply PSLQ again for this improved data in order to check if there is even a smaller relation (with smaller $z_i$). 

Summarizing, the method can be very efficiently used if one knows the set of numbers, by which the result of a 
calculation is finally spanned or if one wants to find a linear combination of such numbers. A highly 
non-trivial example is, e.g., the calculation of the 5--loop $\beta$--function in QCD in 
Ref.~\cite{Luthe:2017ttg}, where these guessing tools were instrumental. 
For a recent survey on these techniques 
(covering not only PSLQ but also the LLL approach) and further applications we refer to~\cite{Acres2021}.
\subsection{Guessing recurrences and differential equations}\label{Sec:GuessREDE}

\vspace*{1mm}
\noindent
In Section~\ref{Sec:largemoment} below we will introduce a method that enables one to compute many moments 
$F(n,\ep)$ in~\eqref{Equ:OtherRepM} or coefficients in~\eqref{Equ:OtherRepPS} for $n=0,1,2,\dots$. 
More precisely, if we write $F(n,\ep)$ in its $\ep$-expansion~\eqref{Equ:EpExpansion},
we will be able to compute the moments of the first $\ep$-coefficients, say $F_{j}(n)$ for $l\leq j\leq r$ with $n=0,1,2,\dots,\mu$ where $\mu$ is large (e.g., $\mu=10^4$). 

Within multi-loop calculations these moments depend linearly also on special constants, such as the 
multiple zeta values~\cite{Blumlein:2009cf}, with rational coefficients. This finally leads to several finite 
sequences, 
$F(0),F(1),\dots,F(\mu)$, of rational numbers. Then given these numbers, one can try to guess a linear recurrence
\begin{equation}\label{Equ:GuessHomRec}
a_0(n)F(n)+a_1(n)F(n+1)+\dots+a_{\lambda}(n)F(n+\lambda)=0
\end{equation}
of order $\lambda$ with polynomial coefficients $a_i(n)\in\QQ[n]$ that contains this finite sequence as solution. Namely, fixing the order $\lambda$ and assuming that the degrees of the polynomials $a_i(n)$ are less than or equal to $\delta$, one searches for the $r=(\delta+1)(\lambda+1)$ unknown coefficients. 
More precisely, by setting $n=0,\dots,r-2$ in~\eqref{Equ:GuessHomRec} and plugging in the rational numbers $F(0),\dots,F(r+\lambda-2)$ one gets $r-1$ equations in $r$ unknowns over the rational numbers which can be solved by linear algebra. In many cases this yields solutions that do not hold for $n\geq r-1$. Thus one usually takes an over-determined system (by more evaluations, say $0\leq n\leq r+100$). In this way one can exclude basically all wrong solutions. Finally, given a found solution one usually checks at many extra points if the recurrence is still valid. This gives further evidence that the guessed recurrence is reliable. 

This tactic implemented, e.g., in the Maple package~\texttt{gfun}~\cite{Salvy:94} or the Mathematica package 
\texttt{GeneratingFunction}~\cite{Mallinger} is surprisingly simple and can be carried out in this naive fashion 
for small examples (i.e., for recurrences of small orders $\lambda$ and small degree bounds $\delta$). For large 
examples within QCD calculations this straightforward procedure utterly fails. Here highly efficient computer algebra technologies, such as homomorphic image calculations and rational/polynomial reconstructions, are essential~\cite{Kauers:08}. Using in addition gcd-calculations to determine recurrences with minimal order, the Mathematica implementation \texttt{Guess.m} by Kauers could be utilized with about $\mu=5000$ moments to guess all the recurrences that determine the massless
unpolarized 3-loop anomalous dimensions and Wilson coefficients in deep-inelastic scattering
\cite{Moch:2004pa,Vogt:2004mw,Vermaseren:2005qc} in Ref.~\cite{Blumlein:2008ouf}, see also 
\cite{Ablinger:2014nga,Ablinger:2017tan,Blumlein:2021enk,Blumlein:2021ryt,Blumlein:2022ndg}.
For even larger problems, 
the highly 
efficient Sage 
implementation 
in~\texttt{ore\_algebra}~\cite{GSAGE} (utilizing among other smart techniques the fast integer arithmetic of 
{\tt Flint}) was instrumental to guess linear recurrences with minimal order.
E.g., for the massive form factor~\cite{Ablinger:2018yae,Blumlein:2019oas} about $\mu=10000$ moments were 
needed to obtain 
recurrences up to order $\lambda=55$ and degree $\delta=1300$.  In the case of a massive operator 
matrix element 8000 moments \cite{Ablinger:2017ptf} could be calculated and difference equations 
were derived for all contributing color and $\zeta$-value structures. Recently, also the 3-loop splitting 
functions~\cite{Ablinger:2017tan}, the anomalous dimensions from off shell operator matrix 
elements~\cite{Ablinger:2017tan,Behring:2019tus,Blumlein:2021enk,Blumlein:2021ryt}, and lately the 
polarized transition matrix element $A_{gq}(N)$~\cite{Behring:2021asx} and the logarithmic contributions 
to the polarized $O(\alpha_s^3)$ asymptotic massive Wilson coefficients~\cite{Blumlein:2021xlc} have been 
derived by guessing the underlying recurrence relations.

Further we note that one can guess in a similar fashion a linear differential equation of the power 
series $f(x)=\sum_{n=0}^{\infty}F(n)x^n$, say
$$a_0(x)f(x)+a_1(x)D_xf(x)+\dots+a_{\lambda}(x)D^{\lambda}_xf(x)=0$$
where $D_x=\frac{d}{dx}$ denotes the differentiation w.r.t.\ $x$. Both, the Mathematica implementation in \texttt{Guess.m} and the Sage implementation in~\cite{GSAGE} cover this extra feature. 

Given such recurrences, one succeeds in many cases to solve the recurrences in terms of special functions that are 
most relevant in QCD calculations. Further details on these solving aspects will be given in the next section.

\section{Solving linear recurrences and differential equations}\label{Sec:LinearSolver}

\vspace*{1mm}
\noindent
As already motivated in Section~\ref{Sec:GuessREDE} above and further emphasized in Sections~\ref{Sec:Summation}--\ref{Sec:largemoment}, one can derive a linear recurrence (linear difference equation) or a linear differential equation which contains the given multi-loop Feynman integral~\eqref{Equ:GeneralForm}
or a given physical expression in terms of such Feynman integrals as a solution. Then a natural strategy is to apply the available toolboxes to compute all solutions of the derived equations that can be represented in terms of certain classes of function spaces that will be introduced in more detail in Section~\ref{Sec:SpecialFunctions}. 
In the case that one finds sufficiently many (linearly independent) solutions one may obtain an alternative 
representation of the physical problem in terms of these solutions. 

In the following we describe different algorithms that can provide solutions of linear difference and differential equations that occur in QCD calculations.

\subsection{Ordinary linear equations}\label{Sec:OrdinaryEquation}

\vspace*{1mm}
\noindent
We start with equations in one variable, i.e., with ordinary linear difference equations of the form
\begin{equation}\label{Equ:ORec}
\sum_{i=0}^{\lambda}a_i(n)\,F(n+i)=r(n)
\end{equation}
and ordinary linear differential equations of the form
\begin{equation}\label{Equ:ODiff}
\sum_{i=0}^{\lambda}a_i(x)D_x^if(x) =r(x)
\end{equation}
where $D_x=\frac{d}{dx}$ denotes the differentiation w.r.t.\ $x$.

\subsubsection{Ordinary linear difference equations}\label{Sec:ScalarRecurrenceSolver}

\vspace*{1mm}
\noindent
The first major contribution for recurrence solving is elaborated in~\cite{Abramov:89a} and is 
substantially improved in~\cite{Hoeij:97}. Given rational functions $a_0(n),\dots,a_{\lambda}(n),r(n)\in\set K(n)$ 
($\set K$ denotes a computable field that contains the rational numbers) it finds all rational solutions $F(n)\in\KK(x)$ of~\eqref{Equ:ORec}. More generally, using the algorithms from~\cite{Petkov:92} and the more efficient version given in~\cite{vanHoeij:99} one can compute all hypergeometric solutions of~\eqref{Equ:ORec}, this means one can compute all solutions that can be written as hypergeometric products
$$F(n)=\prod_{k=l}^nf(k),$$
where $l$ is an integer and $f(k)$ is a rational function in $k$; here $l$ is chosen such that the 
evaluation $f(k)$ for $k\in\set N$ with $k\geq l$ has no pole and is nonzero. In particular, the 
solutions can be described in terms of a product of $\Gamma$-functions, Pochhammer-symbols, 
factorials, binomial coefficients and rational functions. Even more generally, using the algorithms described in~\cite{Abramov:94,Abramov:96} one can search for all d'Alembertian solutions, i.e., all solutions that can be expressed in terms of iterative sums defined over hypergeometric products. Special cases of this class of sums are harmonic sums~\cite{
Blumlein:1998if,Vermaseren:99}, cyclotomic harmonic sums~\cite{Ablinger:2011te}, generalized harmonic 
sums~\cite{Moch:02,Ablinger:2013cf} and finite binomial sums~\cite{Ablinger:2014bra}; infinite binomial sums 
have been 
also 
studied in \cite{Davydychev:2003mv,Weinzierl:2004bn,Ablinger:2014bra}. Further 
details and extra properties of such sums are presented in Section~\ref{Sec:SpecialFunctions}.

Finally, one can search in addition for all Liouvillian solutions~\cite{Singer:99} which cover in addition the interlacing of expressions in terms of iterated sums over hypergeometric products. Basically all these tools have been generalized to the setting of difference fields~\cite{ABPS:21} and rings~\cite{AS:21} (utilizing results from above and~\cite{Karr:81,Bron:00,Schneider:04b,Schneider:05a,Schneider:05b}) that allows one to find such solutions for difference equations~\eqref{Equ:ORec} where the coefficients $a_i(n)$ and the inhomogeneous part $r(n)$ are not just rational functions but can be built again by indefinite nested sums over hypergeometric products.
E.g., using the summation package~\texttt{Sigma}~\cite{Schneider:07a,Schneider:13b,Schneider:21}, that contains this general toolbox, one can compute for the recurrence
\begin{align*}
\big(
1
+{S_1(n)}
+n {S_1(n)}
\big)^2 \big(
3
+2 n
+2 {S_1(n)}
+3 n {S_1(n)}
+n^2 {S_1(n)}
\big)^2 &F(n)\\[-0.1cm]
-(1+n) (3+2 n) {S_1(n)}\big(
3
+2 n
+2 {S_1(n)}
+3 n {S_1(n)}
+n^2 {S_1(n)}
\big)^2 &F(n+1)\\[-0.1cm] 
+(1+n)^2 (2+n)^3 {S_1(n)}\big(
1
+{S_1(n)}
+n {S_1(n)}
\big)  &F(n+2)
=0
\end{align*}
the complete solution set
$$\Big\{c_1\,S_1{n}\,\prod_{l=1}^{n} S_1(l) + c_2S_1(n)^2\,\prod_{l=1}^{n} S_1(l)\mid c_1,c_2\in\set K\Big\};$$
here $S_1(n)=\sum_{k=1}^n\frac1k$ denotes the $n$th harmonic number.
Internally, the recurrence operator is factorized as much as possible into linear factors. Then each extra factor provides one extra linearly independent solution which is constructed by one extra indefinite sum. In other words, finding $\nu$ linear factors (ideally $\nu=\lambda$) yields $\nu$ linearly independent solutions where the 
most complicated solution is built by an iterative nested sum over hypergeometric products of nesting depth $\nu-1$; the particular solution will lead to a nested sum of depth $\nu$. Then a key task is to simplify these sum solutions further such that the nesting depth is minimal; 
further aspects on such simplifications will be given 
in Section~\ref{Sec:SimplificationOfNestedSums}.
We note that all solutions of a linear recurrence can be given in terms
of d'Alembertian solutions if the linear recurrence operator factors
completely into first-order linear factors.
We call such a recurrence also first order factorizing. If this is not
the case, i.e., if only parts of the recurrence can be factored into
linear right-hand factors then it is called non-first order
factorizing.\\ 
The recurrences coming from QCD calculations usually have polynomial coefficients $a_i(n)$ and the right-hand 
side $r(n)$ is either $0$ or is built by indefinite nested sums over hypergeometric products. One of the 
largest homogeneous recurrences ($r(n)=0$) that have been solved with \texttt{Sigma} were of order $\lambda=35$ 
and the degree of the coefficients of $a_i(n)$ was up to $1000$ and the occurring integers required up to 
$1400$ decimals digits; for details see, e.g.,~\cite{Blumlein:2009tj,Blumlein:2008ouf}. The largest inhomogeneous 
recurrences were 
up to order $\lambda=12$ where $r(n)$ may be built up to hundreds of highly nested indefinite nested sums.

In most cases Feynman diagrams or physical expressions in terms of such integrals depend on the dimensional 
parameter $\ep$. In particular, this parameter $\ep$ occurs in the coefficients $a_i(n)$ and the inhomogeneous 
part $r(n)$ of the recurrence~\eqref{Equ:ORec}. In some special cases, the solution $F(n)$ can be given in 
terms of indefinite nested sums over hypergeometric products where $\ep$ occurs inside of the sums and 
products. In such situations, the above methods implemented in~\texttt{Sigma} can find the complete solution in 
$n$ and $\ep$. However, in most instances such a closed form solution does not exist and one seeks for closed 
form solutions of the first coefficients $F_i(n)$ (free of $\ep$) of the 
$\ep$-expansion~\eqref{Equ:EpExpansion}. In order to accomplish this task, one can apply the algorithm 
from~\cite{Blumlein:2010zv} implemented in \texttt{Sigma} in order to constructively decide if the 
coefficients $F_i(n)$ can be represented in terms of nested sums over hypergeometric products. 

The more complicated multi-loop Feynman integrals are considered, the more complicated function spaces arise. Thus
further techniques are extremely desirable that extend the class of indefinite nested sums over hypergeometric 
products. In this regard, one should mention the special case of factorial series 
\cite{NIELSEN:1906,LANDAU:1906,Norlund1924}
solutions of the form 
$f(n)=\sum_{k=0}^{\infty}\tfrac{k!}{(n+k)!}a_k$. Namely, given a linear recurrence in $f(n)$, an operator 
method is described in~\cite{milne-thomson_1932} and further considered in~\cite{Laporta:2000dc}, to provide 
a linear recurrence for the sequence $a_k$. Precisely here one can utilize the recurrence solver of \texttt{Sigma} to decide, if $a_k$ can be written in terms of d'Alembertian solutions.
E.g., for the recurrence
$$(1+n) (2+n) (3+n) f(n)
-(2+n)^2 (3+n) f(1+n)
+(2+n) (3+n) f(2+n)
-f(3+n)
=0$$
one finds the factorial series solution 
$\sum_{k=0}^{\infty}\frac{n! }{(k
	+n
	)!}\sum_{i=0}^k \frac{(-1)^i}{i!}$. Furthermore, M.~Petkov{\v s}ek proposed new ideas in~\cite{Petkov:18} to find solutions of truncated binomial sums: instead of $\frac{k!}{(n+k)!}$ one can choose certain products of binomial coefficients and the upper bound should be integer-linear in $n$. 
\subsubsection{Ordinary linear differential equations}

\vspace*{1mm}
\noindent
In various instances one is interested in a power series solution $f(x)=\sum_{n=0}^{\infty}F(n)x^n$ of a linear differential equation~\eqref{Equ:ODiff}. If the coefficients $a_i(x)$ are rational functions in $x$ and the inhomogeneous part $r(x)$ itself can be given in form of a power series representation, one can utilize holonomic closure properties~\cite{Salvy:94,Mallinger,KP:11} as follows. Plugging the power series ansatz into the differential equation and comparing coefficients w.r.t.\ $x^n$ yield a linear recurrence of the form~\eqref{Equ:ODiff} (with updated $a_i(n)$ and $r(n)$) for the desired coefficients $F(n)$. In a nutshell, one can activate the available recurrence solver introduced in Section~\ref{Sec:ScalarRecurrenceSolver} to compute closed form representations of the coefficients $F(n)$.

Alternatively, there are also direct algorithms available, similarly to the difference equation case, that can solve linear differential equations in terms of rather general classes of special functions. Namely, using the algorithms from~\cite{Abramov:89a} one can find all rational solutions. More generally, using~\cite{Bron:92} and, 
e.g., the improved versions given in~\cite{Hoeij:97,Kauers:13} one can find all hyperexponential functions $f(x)$. In general, the functions can be given in the form $e^{\int_{l}^x h(x)dx}$ for some rational function $h$ and lower bound $l$; special cases are, e.g., rational functions or roots over such functions. More generally, one can use these algorithms to compute all
d'Alembertian~\cite{Abramov:94,Abramov:96}, i.e., all solutions that can be given in terms of iterated integrals over hyperexponential functions. 
Special cases of these integrals, are  harmonic polylogarithms~\cite{Remiddi:1999ew}, cyclotomic 
polylogarithms~\cite{Ablinger:2011te}, generalized multiple polylogarithms~\cite{Moch:02,Ablinger:2013cf} but 
also 
root-valued 
nested integrals~\cite{Ablinger:2014bra}; further details are given in Section~\ref{Sec:SpecialFunctions}.
As for the recurrence case the corresponding differential operator is factorized as much as possible into linear factors. Then each factor yields one extra linearly independent solution by introducing one extra indefinite integration quantifier. These d'Alembertian 
solutions can be computed with the package~\texttt{HarmonicSums}~\cite{Ablinger:2017Mellin}. 
Similarly to the recurrence case we note that all solutions of a linear
differential equation can be given in terms of d'Alembertian solutions
if the differential operator factors completely into first-order linear
factors.
We call such a differential equation also first order factorizing. If
this is not the case, i.e., if only parts of the linear differential
equation can be factored into first-order linear right-factors then it
is called non-first order factorizing.
More generally, also Liouvillian solutions~\cite{Singer:81} can be calculated partially with~\texttt{HarmonicSums} by utilizing Kovacic's algorithm~\cite{Kovacic:86}. For instance, given
\begin{multline*}
(11+20 x) {f\,}'(x)+(1+x) (35+134 x) {f\,}''(x)\\+3 (1+x)^2 (4+37 x) {f\,}^{(3)}(x)+18 x (1+x)^3 {f\,}^{(4)}(x)=0
\end{multline*}
\texttt{HarmonicSums} finds the general solution
\begin{multline*}
\Big\{c_1+c_2 \int_0^x \tfrac{1}{1+\tau _1} \, d\tau _1+c_3 \int _0^x\int _0^{\tau _1}\tfrac{\sqrt[3]{1+\sqrt{1+\tau
			_2}}}{\left(1+\tau _1\right) \left(1+\tau _2\right)}d\tau _2d\tau _1\\
+c_4 \int _0^x\int _0^{\tau _1}\tfrac{\sqrt[3]{1-\sqrt{1+\tau _2}}}{\left(1+\tau _1\right) 
\left(1+\tau _2\right)}d\tau _2d\tau_1\mid c_1,c_2,c_3,c_4\in\set K\Big\},
\end{multline*}				
where $\sqrt{1+x}$ is hyperexponential and $\sqrt[3]{1-\sqrt{1+x}}$ is algebraic over a field generated by $x$ and $\sqrt{1+x}$.
More generally, in~\cite{Singer:81} an algorithm has been described that finds all Liouvillian solutions of a 
homogeneous linear differential equations, i.e., all solutions that can be given by iterated integrals over 
hyperexponential function and functions that are algebraic over the extension below. More generally, an algorithm has been proposed in~\cite{Singer:91} that can find all Liouvillian solutions of linear differential equations whose coefficients are given in terms of functions that are Liouvillian. In some sense, this highly general solver can be considered as the continuous version of the recurrence solver~\cite{ABPS:21} implemented within the package~\texttt{Sigma}.

As already emphasized in Section~\ref{Sec:ScalarRecurrenceSolver},  also the dimensional parameter $\ep$ appears in the coefficients $a_i(x)$ and the inhomogeneous part $r(x)$ of the linear differential equation~\eqref{Equ:ODiff} when one deals with Feynman integrals. In some special cases one can use the above algorithms directly where $\ep$ may arise inside of d'Alembertian and Liouvillian solutions. However, similarly to the recurrence case, this approach usually does not work and one aims at finding closed forms of the first coefficients of the $\ep$-expansion 
$$f(x,\ep)=f_l(x)\ep^l+f_{l+1}(x)\,\ep^{l+1}+\dots+f_r(x)\,\ep^r+O(\ep^{r+1}).$$ 
In this regard, the package \texttt{HarmonicSums} can decide constructively if the first coefficients $f_i(x)$ 
(free of $\ep$) can be given in terms iterated integrals over hyperexponential functions; for the underlying 
algorithm we refer to~\cite{Ablinger:21} which is based on ideas given in~\cite{Blumlein:2010zv}.

By looking at more and more complicated Feynman integrals also the 
class of Liouvillian solutions is not sufficient. For second order 
linear differential equations van Hoeij proposed algorithms 
in~\cite{IVH} that can find hypergeometric series solutions 
(${}_pF_q$'s) in terms of certain rational function arguments. These 
advanced tools turned out to be instrumental to deal with the 
$\rho$-parameter in~\cite{Ablinger:2017bjx,Blumlein:2018aeq} and related quantities.

\subsection{Partial linear equations}

\vspace*{1mm}
\noindent
Solving partial linear difference and differential equations is a hard problem. It has been shown in~\cite{AP:12} based on~\cite{Hilbert10} that already the task to solve such equations in terms of polynomial solutions is an unsolvable problem. 
Recently, new methods have been introduced in~\cite{KS:10,KS:11} that enable one to search for (not necessarily all) rational solutions of partial linear difference equations of the form
\begin{equation}\label{Equ:PLRecE}
\sum_{(s_1,\dots,s_r)\in S} a_{(s_1,\dots,s_r)}(n_1,\dots,n_r)F(n_1+s_1,\dots,n_r+s_r)=0,
\end{equation}
where the coefficients $a_{(s_1,\dots,s_r)}$ are rational functions in the variables $n_1,\dots,n_r$ and $S\subset\set Z^r$ is a finite set. 
In~\cite{Blumlein:2021hbq}
further ideas coming from Section~\ref{Sec:ScalarRecurrenceSolver} have been incorporated to hunt also for solutions in terms of a given set of nested sums.  E.g., suppose that we are given the partial linear difference equation
\begin{multline*}
-(n+1)^2
\left(k+n^2+2\right)
\left(4 k^2-3 k n^2+5 k
n+12 k-2 n^3-2 n^2+8
n+8\right) F(n,k+1)\\
+(n+1)^2
\left(k+n^2+3\right)
\left(2 k^2-2 k n^2+2 k n+6
k-n^3-n^2+4 n+4\right)F(n,k+2)\\
+(n+1)^2 (k+n+1)
\left(2 k-n^2+n+4\right)
\left(k+n^2+1\right)
F(n,k)\\
-(k+1) n^2 (n+2)^2  
\left(k+n^2+2 n+2\right)F(n+1,k)\\
+k n^2 (n+2)^2
\left(k+n^2+2 n+3\right)
F(n+1,k+1)=0
\end{multline*}
and the set $W=\{S_1(k),S_1(n+k),S_{2,1}(n+k)\}$ in terms of the harmonic numbers and the harmonic sum 
$S_{2,1}(n)=\sum_{k=1}^n\frac{S_1(k)}{k^2}$; compare Section~\ref{Sec:SpecialFunctions}. Then fixing the total degree bound $5$ or the arising objects in the numerator, one can compute with the package~\texttt{SolvePLDE} introduced 
in~\cite{Blumlein:2021hbq}
the $37$ solutions $\frac{p}{(1 + n)^2 (1 + k + n^2)}$ where $p$ is taken from the set
\small
\begin{align*}
\big\{&1
+\frac{1}{2} n S_1({k+n})
,k,n,k n,k n^2,k n^3,k n^4,k S_1({n}),k n S_1({n}),k n^2 S_1({n}),k n^3 S_1({n}),k S_1({n})^2,\\
&k n S_1({n})^2,k n^2 S_1({n})^2,k S_1({n})^3,k n S_1({n})^3,k S_1({n})^4,k S_{2,1}({n}),k n S_{2,1}({n}),k n^2 S_{2,1}({n}),k n^3 S_{2,1}({n}),\\
&k S_1({n}) S_{2,1}({n}),k n S_1({n}) S_{2,1}({n}),k n^2 S_1({n}) S_{2,1}({n}),k S_1({n})^2 S_{2,1}({n}),k n S_1({n})^2 S_{2,1}({n}),\\
&k S_1({n})^3 S_{2,1}({n}),k S_{2,1}({n})^2,k n S_{2,1}({n})^2,k n^2 S_{2,1}({n})^2,k S_1({n}) S_{2,1}({n})^2,k n S_1({n}) S_{2,1}({n})^2,\\
&k S_1({n})^2 S_{2,1}({n})^2,k S_{2,1}({n})^3,k n S_{2,1}({n})^3,k S_1({n}) S_{2,1}({n})^3,k S_{2,1}({n})^4\big\}.
\end{align*}
\normalsize

In particular, the method for scalar linear difference equations in~\cite{Blumlein:2009tj} has been carried 
over 
in this new package to search also for closed form solutions of the first coefficients of an $\ep$-expansion.

We emphasize that this new package enables one also to attack partial linear differential equations and to find solutions in its multivariate power series expansion $f(x_1,\dots,x_r)=\sum_{(n_1,\dots,n_r)\in\set N^r}F(n_1,\dots,n_r)x_1^{n_1}\dots x_r^{n_r}$. Namely by plugging the power series ansatz into the partial differential equation and comparing coefficients w.r.t.\ $x_1^{n_1}\dots x_r^{n_r}$ produce a partial linear difference equation of the form~\eqref{Equ:PLRecE}. Thus one can utilize the tools described above to search for closed form representations of $F(n_1,\dots,n_r)$. In Section~\ref{Sec:DirectCoupledSolver} this tactic will be refined further to find solutions for certain classes of coupled systems of partial linear differential equations.

\section{Solving coupled systems of linear differential equations}\label{Sec:SolveSystems}

\vspace*{1mm}
\noindent
In order to solve open problems at the forefront of elementary particle physics, millions of 
complicated Feynman 
integrals have to be tackled. 
As a preprocessing step one often applies integration-by-parts (IBP) 
methods~\cite{Chetyrkin:1981qh,Laporta:2001dd,Marquard:2021spf}\footnote{Here the method of syzygies 
\cite{HILBERT,Gluza:2010ws} from computational algebraic geometry helps to reduce the number of contributing 
scalar products.} 
that crunch these integrals to a few hundred (or thousand) so-called master integrals; for a recent survey, 
possible refinements and applications see, e.g.,~\cite{Marquard:2021spf,Vermaseren2021,Kotikov2021}. Then the main 
task is to simplify only these master integrals to expressions in terms of special functions and to assemble the original problem with these sub-results. 
Most of these master integrals $f_i(x,\ep)$ can be determined as solutions of coupled systems of linear differential equations.
For single-variate systems they are of the form
\begin{eqnarray}
\label{eq:DEQ1}
\hspace*{2cm}\Dx 
\left(
\begin{smallmatrix}
f_1(x,\ep)\\ \vdots \\ f_{\lambda}(x,\ep)\end{smallmatrix}\right) 
= A
\left(\begin{smallmatrix} f_1(x,\ep)\\ \vdots \\ f_{\lambda}(x,\ep)\end{smallmatrix}\right) 
+ \left(\begin{smallmatrix} g_1(x,\ep)\\ \vdots \\ g_{\lambda}(x,\ep)\end{smallmatrix}\right),
\end{eqnarray}
with $A$ being a $\lambda\times\lambda$ matrix with entries from $\KK(x,\ep)$ where the right-hand sides are given in terms of simpler master integrals. They are either determined by other coupled systems or have to be tackled by tools presented, e.g., in Sections~\ref{Sec:Summation} and~\ref{Sec:Integration}.
Here we elaborate the most relevant approaches. Before one considers to solve such systems, one may also analyze them further as exemplified in~\cite{Dreyfus2021} in order gain further insight or to find further relations among them. 

\subsection{Uncoupling algorithm and scalar solvers}

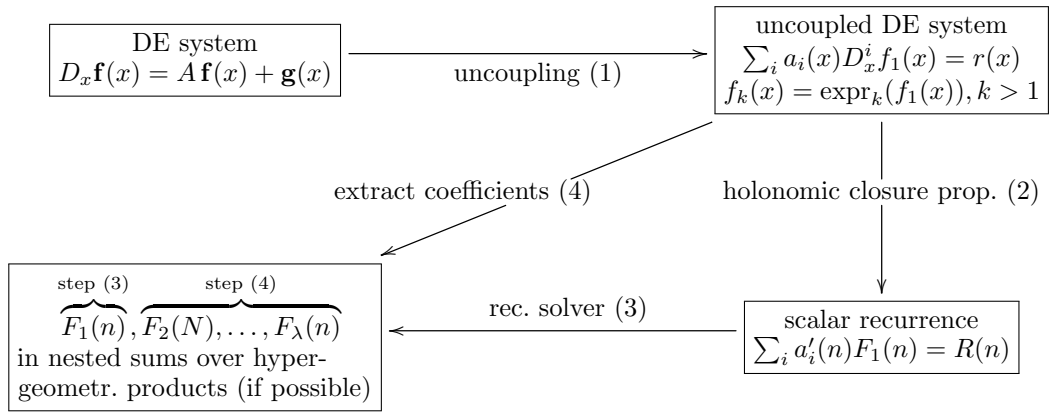
\begin{figure}
	\centering
	\footnotesize
	$$\hspace*{-0.2cm}\xymatrix@!R=1cm@C2cm{
		\boxed{\txt{DE system\\
				$D_x\vect{f}(x)=A\,\vect{f}(x)+\vect{g}(x)$}}{\ar[rr]_{\txt{uncoupling (1)}}}&&
		{\boxed{\txt{uncoupled DE system\\
					$\sum_i a_i(x)D_x^i f_1(x)=r(x)$\\
					${\bf}f_k(x)=\text{expr}_k(f_1(x)), k>1$}}{\ar[dd]|{\txt{holonomic closure prop.\ (2)}}}
			{\ar[ddll]|{\txt{extract coefficients (4)\hspace*{2cm}}}}}\\
		\\
		{\boxed{\txt{
					$\overbrace{\text{$F_1(n)$}}^{\text{step (3)}},\overbrace{\text{$F_{2}(N),\dots, F_{\lambda}(n)$}}^{\text{step (4)}}$\\
					in nested sums over hyper-\hspace*{0.6cm}\\ 
					geometr.\ products (if possible)\hspace*{0.cm}}}}
		&&{\boxed{\txt{scalar recurrence\\
					$\sum_i a'_i(n)F_1(n)=R(n)$}}{\ar[ll]_{\txt{\hspace*{0.8cm}rec.\ solver (3)}}}}\\
	}$$
	\normalsize
	\caption{\label{fig:uncouplingtactic}{Solving systems.}}
\end{figure}

\vspace*{1mm}
\noindent
In the last years a general toolbox has been elaborated that finds all solutions that can be given in terms of iterated integrals (or sums) as follows.
By uncoupling algorithms~\cite{Zuercher:94,BCP13} available, e.g., in the package~\texttt{OreSys}~\cite{ORESYS}, one first decouples the system~\eqref{eq:DEQ1} to a scalar linear differential equation in one of the unknowns. Using the differential equation solver in~\texttt{HarmonicSums}~\cite{Ablinger:2017Mellin} (based on~\cite{Singer:81,Kovacic:86,Abramov:94}) one finds, whenever possible, a closed form representations of the unknown functions $f_1,\dots,f_{\lambda}$ in terms of d'Alembertian (and partially of Liouvillian) solutions.
Based on this strategy we recalculated the 2-loop form factors~\cite{Ablinger:2017hst} and obtained first 
results for the 3-loop 
case~\cite{Ablinger:2018zwz}. Another fruitful 
approach~\cite{Ablinger:2015tua} is 
based on recurrence solving. Here one assumes that the arising Feynman integrals can be 
given in the power series representations
\begin{equation}\label{Equ:GenerateFu}
f_i(x)=\sum_{n=0}^{\infty}F_i(n) x^n,\quad i=1,\dots,\lambda.
\end{equation}
Then the machinery proceeds as summarized in Fig.~\ref{fig:uncouplingtactic}.
After the decoupling of the system (step~1) one takes the scalar differential equation of $f_1(x)$ and calculates by means of holonomic closure properties~\cite{KP:11} a scalar linear recurrence of $F_1(n)$ (step~2). 
Activating the recurrence solver of~\texttt{Sigma} (based 
on~\cite{Abramov:89a,Petkov:92,vanHoeij:99,Abramov:94,Schneider:05a,Blumlein:2010zv}) one can decide 
algorithmically in 
step~3 if $F_1(n)$ can be represented in terms of indefinite nested sums. If yes, one plugs this representation into the decoupled system and gets closed forms of the remaining $F_2(n),\dots,F_{\lambda}(n)$ in step~(4). 
For advanced QCD-calculations see, 
e.g.,~\cite{Ablinger:2014nga,Behring:2015zaa,Behring:2015roa,Ablinger:2015tua,
Behring:2016hpa,Ablinger:2017tan,Blumlein:2012vq,Ablinger:2014lka}

\subsection{Direct solver}\label{Sec:DirectCoupledSolver}

\vspace*{1mm}
\noindent
So far there are only few algorithms available that can compute directly (i.e., without uncoupling) 
the desired set of solutions of a given coupled system of the form~\eqref{eq:DEQ1}. 
For instance, with the algorithms in~\cite{BR:12} one can find all hyperexponential solutions of (higher order) coupled systems but so far no algorithms are available to compute all d'Alembertian or Liouvillian solutions. First steps have been elaborated in~\cite{MS:18} within the general difference field setting.

However, in various instances it has been observed in~\cite{Henn:2013pwa} that the arising coupled systems of the form~\eqref{eq:DEQ1} can be transformed to a system of the form
\begin{equation}\label{Equ:SpecialTransformation}
\Dx\vect{\tilde{f}}(x,\ep)=\ep \tilde{A}(x)\vect{\tilde{f}}(x,\ep)+\vect{\tilde{g}}(x,\ep)
\end{equation}
for a matrix $\tilde{A}(x)$ which is free of $\ep$ and where $\vect{\tilde{f}}$ and 
$\vect{\tilde{g}}$ are defined by the multiplication of an invertible matrix $T$ with 
$\vect{f}=(f_1,\dots,f_{\lambda})$ and $\vect{g}=(g_1,\dots,g_{\lambda})$, respectively. Under the 
assumption that such a transformed system exists, algorithms are available to 
compute such a transformation matrix $T$. Furthermore, methods exist, 
cf.~\cite{Lee:2014ioa,Gituliar:2017vzm,Prausa:2017ltv,Meyer:2017joq}
that can hunt for such a basis transformation for the multivariate case, i.e., for systems 
of partial linear differential equations. Given such a transformed system one obtains the 
benefit that one can read off the coefficients of the $\ep$-expansion in terms of indefinite 
nested integrals. However, such a transformation does not hold for more complicated systems. 

As elaborated in~\cite{Blumlein:2021hbq},
there are other special cases that enable one to solve coupled systems of partial linear differential equations if one considers the solution as a multivariate power series solution where the coefficients satisfy a nicely coupled system of linear difference equations. 
For instance, take the coupled partial system
\begin{align*}
(x-1) y D_{x y}f(x,y)+(x (2 \varepsilon +\tfrac{7}{2})-\varepsilon +1)D_x f(x,y)+(x-1) x D_x^2f(x,y)&\\
+y (2 \varepsilon +1)D_yf(x,y)+\tfrac{3}{2} (2 \varepsilon +1)f(x,y)&=0,\\
x (y-1) D_{x y}f(x,y)+x (4-\varepsilon ) D_xf(x,y)+ (y (\tfrac{13}{2}-\varepsilon
)-\varepsilon +1)D_yf(x,y)&\\
+(y-1) y D_y^2f(x,y)+\tfrac{3 (4-\varepsilon )}{2}f(x,y)&=0.
\end{align*}
Then writing $f(x,y)$ as a multivariate power series $f(x,y)=\sum_{n,m=0}^{\infty}F(n,m)x^ny^m$ one obtains by coefficient comparison w.r.t.\ $x^nx^m$ the coupled system of homogeneous first-order difference equations
\begin{align*}
\frac{3}{2} (2 \varepsilon +1) F(n,m)-n (\varepsilon -1) F(n+1,m)&=0,\\
-\frac{3}{2} (\varepsilon -4) F(n,m)-m (\varepsilon -1)F(n,m+1)&=0.
\end{align*}
Given such a first-order homogeneous system, it follows that its solution can be expressed in terms of hypergeometric products. Namely, using the algorithm given 
in~\cite[Sec.~4.1]{Blumlein:2021hbq}
(which is a simplified version of the algorithm given in~\cite{AP:02}) and implemented in the 
package~\texttt{HypSeries} one obtains the solution
\begin{align*}
F(n,m)=&\big(
\prod_{i=1}^n \frac{(1+2 i) (3
	+i
	-\varepsilon 
	)}{2 i (-2
	+i
	+\varepsilon 
	)}\big) \prod_{i=1}^m \frac{(1
	+2 i
	+2 n
	) (i
	+2 \varepsilon 
	)}{2 i (-2
	+i
	+n
	+\varepsilon 
	)}\\
=&\frac{\big(
	\frac{3}{2}\big)_{m
		+n
	} (4-\varepsilon )_n (1+2 \varepsilon )_m}{m! n! (-1+\varepsilon )_{m
		+n
}}
\end{align*}
in terms of hypergeometric products or equivalently in terms of factorial and Pochhammer symbols. As a consequence the derived solution of the original coupled system of differential equations can be given in the form
$$F(x,y)=\sum_{n,m=0}^{\infty}\frac{\big(
	\frac{3}{2}\big)_{m
		+n
	} (4-\varepsilon )_n (1+2 \varepsilon )_m}{m! n! (-1+\varepsilon )_{m
		+n
}}x^ny^m.$$
Using this sum representation one can deploy the summation tools in Section~\ref{Sec:Summation} to calculate the first coefficients of its $\ep$-expansion.
We note that these solutions (coming from homogeneous first-order difference systems) are closely 
related to special functions that are introduced in the next section.

\section{Transformation to special integral and sum representations}
\label{Sec:Transformation}

\vspace*{1mm}
\noindent
In the simplest cases, the integrands of Feynman integrals~\eqref{Equ:GeneralForm} exhibit Euler Beta-function 
structures and by clever rewriting, cf.\ also e.g. \cite{HAMBERG}, the integral can be rewritten in terms of 
hypergeometric functions 
and their generalization~\cite{KLEIN:1893,BAILEY:1935,SLATER:1966,AAR}.
More precisely, one may rewrite simple Feynman integrals in terms of the following hierarchy of $_{p+1}F_p$ functions, the first of
which read
\begin{align*}
B(a_1,a_2)           &= \int_0^1 dt~t^{a_1-1} (1-t)^{a_2-1}
\\
\label{eq:2F1}
_2F_1(a_1,a_2,b_1;x) &= 
\frac{\Gamma(b_1)}{\Gamma(a_2)\Gamma(b_1-a_2)} \int_0^1 dt~t^{a_2-1} 
(1-t)^{b_1-a_2-1} (1-tx)^{-a_1}
\\
_3F_2(a_1,a_2,b_1;x) &= \frac{\Gamma(b_2)}{\Gamma(a_3) \Gamma(b_2-a_3)} \int_0^1 dt~t^{a_3-1} 
(1-t)^{-a_3+b_2-1}~_2F_1(a_1,a_2,b_1;tx).
\end{align*}
Here the parameters $a_i, b_i$ are such, that the corresponding integrals exists, \cite{AAR}.\footnote{One may 
then perform corresponding analytic continuations, cf.~\cite{WW,KF}.} 
We note that computer algebra can be used non-trivially
to explore further properties on these special functions. E.g., using symbolic summation (see also Section~\ref{Sec:Summation}) one can compute all arising
contiguous relations of a finite set of sums~\cite{Paule2021}. More generally, it is possible to represent 
Feynman integrals by Mellin--Barnes \cite{MELLIN,BARNES1,BARNES2} representations \cite{Blumlein:2010zv}.
As well--known, the Mellin--Barnes representations are also used for hypergeometric functions and their 
generalizations, originally in terms of Pochhammer Umlauf--integrals.

At 3-loop order, also Appell functions 
\cite{APPEL1,APPEL2,KAMPE1,EXTON1,EXTON2,SCHLOSSER,SRIKARL,Kalmykov2021,Blumlein:2021hbq}
and their generalizations arise; see, e.g.,~\cite{Anastasiou:1999ui,Ablinger:2012qm,Ablinger:2015tua}. For 
instance, 
the $F_1$ 
function has the integral representation
\begin{align*}
F_1(a,b_1,b_2,c;x,y) &= \frac{\Gamma(c)}{
	\Gamma(a) \Gamma(c-a)} \int_0^1 dt~t^{a-1} (1-t)^{c-a-1} (1-xt)^{-b_1} (1-yt)^{-b_2},\\
&\hspace*{9cm} {\sf Re}(c) > {\sf Re}(a) > 0.
\end{align*}
When one succeeds in detecting that the the given Feynman integrals can be rewritten in integral representations that can be connected to $_{p+1}F_p$ or Appell-like functions, one can utilize the essential property that all the  $_{p+1}F_p$ functions have a single infinite sum representation, while the Appell-functions
are represented by two infinite sums. For instance, we get
$$F_1(a;b_1,b_2;c;x,y) = \sum_{m=0}^\infty \sum_{n=0}^\infty 
\frac{(a)_{m+n} (b_1)_m (b_2)_n}{m! n! (c)_{m+n}} x^m y^n.$$
Similarly, there are also other classes of higher transcendental 
functions, 
which obey multi-sum representations \cite{EXTON1,EXTON2,SRIKARL,Blumlein:2021hbq}.
Up to the level of massless and single mass two-loop integrals, cf.~\cite{Bierenbaum:2007qe} 
and in some cases in the three loop case, cf.~\cite{Ablinger:2015tua}, these representations are 
usually sufficient. For more complicated integrand-structures, however, one has to apply other 
techniques. Applying successively Newton's binomial theorem and Mellin--Barnes 
\cite{MELLIN,BARNES1,BARNES2} decompositions on 
the integrand, implemented in different packages 
\cite{Czakon:2005rk,Smirnov:2009up,Gluza:2007rt,PhysRevLett.127.151601},
enables one to carry out all integrals by 
introducing Mellin--Barnes integrals. Finally, carrying out the remaining Mellin--Barnes 
integrals with the residue theorem yields definite multiple sums 
\begin{equation}
\label{Equ:SumRep} \sum_{k_1=1}^{L_1(n)}\dots\sum_{k_v=1}^{L_v(n,k_1,\dots, 
k_{v-1})} f(n,k_1,\dots,k_v). 
\end{equation} 
Here the upper bounds 
$L_1(n),\dots,L_{v}(n,k_1,\dots,k_{v-1})$ are integer linear (i.e., linear combinations of the 
variables over the integers) in the dependent parameters or $\infty$, and $f$ is hypergeometric 
in $n$ and the summation variables $k_i$. For further details on this rewriting process we 
refer, e.g., to~\cite{Blumlein:2010zv,Weinzierl:22}.

In physical applications the dimensional parameter $\varepsilon$ arises in the parameters $a_i, 
b_i,c_i, ...$ of the $_{p+1}F_p$ and Appell-type representations.
Moreover, the hypergeometric 
summand $f$ in~\eqref{Equ:SumRep} may also depend on $\ep$. In all these cases one seeks for an 
$\ep$-expansion where the coefficients are represented in sum representations that are as 
simple as possible. In order to accomplish this task, highly general summation methods 
introduced in Section~\ref{Sec:Summation} can be applied.
\section{Symbolic summation}\label{Sec:Summation}

\vspace*{1mm}
\noindent
Following the strategy sketched in Section~\ref{Sec:Transformation} one ends up at (thousands or even 
millions) of definite multiple sums where the summand is built by hypergeometric products and indefinite 
nested sums, like harmonic sums~\cite{Blumlein:1998if,Vermaseren:99}, cyclotomic harmonic 
sums~\cite{Ablinger:2011te}, 
generalized 
harmonic sums~\cite{Moch:02,Ablinger:2013cf}; these sums may pop up in particular if one expands the summand 
w.r.t.\ the 
$\ep$-parameter (i.e., if one applies the differential operator w.r.t.\ $\ep$ to the hypergeometric products; for a detailed description see, 
e.g.,~\cite{Blumlein:2021hbq}).
Producing such sum representations without making the original problem more complicated is highly non-trivial. However, if one succeeds in getting an appropriate sum representation, one can apply various symbolic summation algorithms to simplify these sums.

\subsection{Simplification of indefinite nested sums}\label{Sec:SimplificationOfNestedSums}

\vspace*{1mm}
\noindent
The simplification of indefinite nested sums defined over hypergeometric products started with Gosper's and Karr's summation algorithms~\cite{Gosper:78,Karr:81} and has been enhanced significantly within the last 20 years to a strong summation machinery based on difference field and ring theories~\cite{Bron:00,Schneider:01,DR1a,DR1,DR3,DR3b}. Using our summation package \texttt{Sigma}~\cite{Schneider:07a,Schneider:13b} it is now possible to design completely automatically appropriate difference rings in which one can represent such indefinite nested sums fulfilling various optimality criteria: e.g., the number of nested summation quantifiers or the degrees in the denominators are minimized; see~\cite{Schneider:07d,Schneider:08c,Schneider:10a,Schneider:15}. Furthermore, employing our contributions to a refined Galois theory of difference rings~\cite{DR2} (see also~\cite{Singer:99,Kauers:08b,Singer:08,Schneider:10c}), the used sums do not admit any algebraic relations. As a consequence, one obtains canonical (unique) product-sum representations~\cite{Schneider:21}.

Furthermore, these algorithms can be accompanied with quasi-shuffle relations~\cite{Hoffman,
Blumlein:2003gb,
Blumlein:2009ta,
Blumlein:2009fz,AS:18} for the discovery of such relations in a very efficient way; for further details we refer to Section~\ref{Sec:SpecialFunctions}.

\subsection{The WZ-summation approach}

\vspace*{1mm}
\noindent
The treatment of single nested definite hypergeometric sums started with Zeilberger's creative telescoping paradigm~\cite{Zeilberger:91,PauleSchorn:95,Paule:95,CK:12,BostanDumontSalvy2016}  and has been enhanced to multi-summation with the WZ-summation approach due to~\cite{Wilf:92} and its refinements given, e.g., in~\cite{Wegschaider,LPR:02,AZ:06}. Given a multiple sum $F(n,\ep)$ over a hypergeometric summand, like on the left-hand side of\footnote{For $a\in\ZZ\setminus\{0\}$ we define the generalized harmonic numbers 
$S_a(n)=\sum_{k=1}^n\frac{(\textrm{sign}(a))^k}{k^{|a|}}$.}
\begin{multline}\label{Equ:SimpleTripleSum}
\sum_{j=0}^{n-2} \sum_{r=0}^{j+1} \sum_{s=0}^{n-j+r-2} \frac{(-1)^r (n-j-2)!
\binom{j+1}{r} r!}{(n-j+r)!}\frac{(-1)^s \binom{n-j+r-2}{s}}{(n-s)(s+1)}\\
=\frac{((-1)^n-1)(n^2+n+1)}{n^2 (n+1)^3}+\frac{S_1(n)}{(n+1)^2}-\frac{S_2(n)+2
S_{-2}(n)}{n+1},
\end{multline}
one can search for a recurrence/difference equation of order $\lambda$ of the form
$$\sum_{i=0}^{\lambda}a_i(n,\ep)\,F(n+i,\ep)=r(n,\ep)$$
with polynomials $a_i(n,\ep)$ in $n,\ep$ and $r(n,\ep)$ being an expression in terms of multiple sums of simpler type than $F(n,\ep)$. By further tricks one can compute even a homogeneous recurrence.
With the package \texttt{MultiSum}~\cite{Wegschaider} one obtains, for instance, for the triple sum in~\eqref{Equ:SimpleTripleSum} a homogeneous linear recurrence with polynomial coefficients in $n$ of order $\lambda=4$ in about 2 days.

Given this recurrence, one can utilize algorithms from~\cite{Abramov:89a,Petkov:92,vanHoeij:99,Singer:99,Schneider:05a,ABPS:21} (see Section~\ref{Sec:ScalarRecurrenceSolver}) encoded in our package~\texttt{Sigma} that find all d'Alembertian solutions, i.e., all solutions  that can be expressed in terms of indefinite nested sums 
defined over hypergeometric products.
More precisely, \texttt{Sigma} computes 4 linearly independent solutions (i.e., their linear span generates all solutions) where 
\begin{equation}\label{Equ:SolutionNotSimplified}
\tfrac{-1}{4 (1+n)^2}\sum_{i=3}^n
\sum_{j=3}^i \tfrac{(
	8-24 j+11 j^2+3 j^3-3 j^4+j^5) 
}{(-2+j)^2 (-1+j)^2 j^2 (1+j)}\sum_{k=1}^j \tfrac{(-1)^k (-2+k)^2 (
	9-86 k+229 k^2-156 k^3-26 k^4+64 k^5-26 k^6+4 k^7)}{(
	36-20 k-26 k^2+25 k^3-8 k^4+k^5
	)(8-24 k+11 k^2+3 k^3-3 k^4+k^5)}
\end{equation}
is the most complicated sum solution. Finally, with four initial values of the triple sum one finds an alternative representation of it in terms of indefinite nested sums.

In general, these are highly nested, and the summands might consist of ugly polynomials in the denominator (like in~\eqref{Equ:SolutionNotSimplified}) that do not factorize nicely. However, employing our sophisticated difference ring algorithms introduced in Section~\ref{Sec:SimplificationOfNestedSums}, one can simplify the found representation further and obtains the right-hand side in~\eqref{Equ:SimpleTripleSum}. In total, the solving and simplification steps need around 10 seconds.

Summarizing, combining the WZ-approach (recurrence finding) and solving tools, one obtains a summation machinery that can transform a definite nested sum to expressions in terms of indefinite nested sums. 
When the input sum depends furthermore on the dimensional parameter $\ep$, this machinery has been generalized 
in~\cite{Blumlein:2010zv} to determine the coefficients of the $\ep$-expansion of~\eqref{Equ:EpExpansion} 
whenever they are expressible in terms of indefinite nested sums defined over hypergeometric products.
This toolbox is very general, but has a substantial drawback: it reaches already with such simple sums like in~\eqref{Equ:SimpleTripleSum} its limit. With the  difference ring approach described next, this situation can be improved substantially.
\subsection{The difference ring approach}

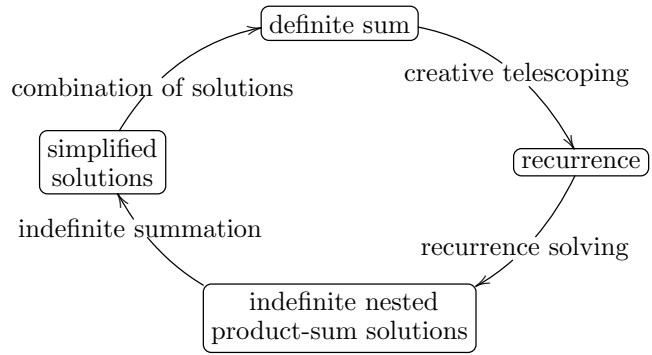
\begin{figure}
	\centering
	\footnotesize
	$$\xymatrix@R=1.2cm@C=0.5cm{
		&*+[F-:<3pt>]{\txt{definite sum}} \ar @/^1.5pc/[rd]|>>>>>>>>>>{\txt{creative telescoping}}&\\
		*+[F-:<3pt>]{\txt{simplified\\ solutions}}\ar @/^1.5pc/[ru]|>>>>>>>>>>>{\txt{combination of
				solutions}} &&*+[F-:<3pt>]{\txt{recurrence}}\ar @/^1.5pc/[ld]|<<<<<<<<<<{\txt{recurrence solving}}\\
		&*+[F-:<3pt>]{\txt{ 
				indefinite nested\\ 
				product-sum solutions}}\ar @/^1.5pc/[lu]|<<<<<<<<<{\txt{indefinite
				summation}}& }$$
	\normalsize
	\caption{\SigmaP's summation spiral\label{fig:Spiral}; see~\cite{Schneider:04c}.}
\end{figure}

\vspace*{1mm}
\noindent
With the difference ring and field theories worked out in~\cite{Karr:81,Schneider:01,Schneider:08c,Schneider:15,DR1,DR2} one can simplify not only indefinite nested sums, but one can also apply Zeilberger's creative telescoping paradigm~\cite{Zeilberger:91}. This means that one can try to compute 
a linear recurrence of order $\lambda$ for a definite sum, say $S(n)=\sum_{k=0}^nf(n,k)$, where $f(n,k)$ is given in terms of indefinite nested sums defined over hypergeometric products w.r.t.\ the summation variable $k$. Given such a recurrence, one can solve it in terms of indefinite nested sums defined over hypergeometric products by the algorithms given in Section~\ref{Sec:ScalarRecurrenceSolver}. If one succeeds in combining the solutions accordingly (matching $\lambda$ initial values), one obtains an alternative representation of $S(n)$. If this expression itself is summed over $n$, one can repeat this process w.r.t.\ another variable (over which one may sum later again). In a nutshell, one can apply the summation spiral illustrated in Fig.~\ref{fig:Spiral} iteratively with the goal to transform a given multi-sum from inside to outside to a representation purely in terms of indefinite nested sums. 

This interplay has been automated in the package 
\texttt{EvaluateMultiSums}~\cite{Schneider:13b} based on \texttt{Sigma}'s difference ring algorithms and produces the right-hand side in~\eqref{Equ:SimpleTripleSum} in about 70~seconds. 
\begin{figure}
	\centering
¸	\includegraphics[width=1.6cm,height=1.6cm]{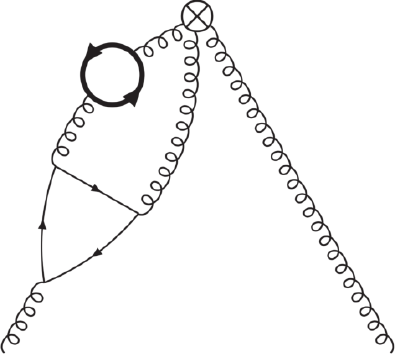}
	
	\vspace*{-0.4cm}
	
	\caption{\label{fig:2Mass3Loop}2-mass 3-loop diagram.}
	
	\vspace*{-0.3cm}
	
\end{figure}
If a sum depends also on the dimensional parameter $\ep$, one can first expand the summand of the multi-sum w.r.t.\ $\ep$ and can apply afterwards the summation quantifiers to each of the coefficients being free of $\ep$.
A clear drawback of this approach is that the summands blow up when higher $\ep$-orders are calculated. Nevertheless, 
the pure difference ring approach produces simplifications that currently no other toolbox can achieve.
E.g., while treating the 2-mass 3-loop integral given in Fig.~\ref{fig:2Mass3Loop}
triple and quadruple sums between 0.4 to 1.6 GB of memory arose in 
~\cite{Ablinger:2018brx} that could be simplified to expressions in 
terms of binomial sums using 8.4 MB of memory only. Further 
challenging calculations based on the difference ring/field approach 
can be found, e.g., 
in~\cite{Ablinger:2015tua,Ablinger:2018brx,Ablinger:2020snj,Ablinger:2019gpu}.
\subsection{The holonomic-difference ring approach}

\vspace*{1mm}
\noindent
Another prominent branch of symbolic summation is the holonomic system approach which has been introduced in~\cite{Zeilberger:90a} and pushed further, e.g., in~\cite{Chyzak:00,Koutschan:13} to determine recurrence relations. Here the summand of a definite sum is described by a system of homogeneous recurrences with polynomial
coefficients. Then given such a system, one can try to compute a linear recurrence system by introducing the 
next definite summation quantifier. Applying these algorithms iteratively from inside to outside yields a 
linear recurrence in $n$ for the input sum. However, the underlying recurrence systems may grow heavily and the 
holonomic approach usually fails due to time and memory limitations. In~\cite{Schneider:05d} a hybrid strategy 
has been introduced and developed further in~\cite{Ablinger:2012ph,DRHolonomic}
that brings the holonomic and difference ring/field approach under a common umbrella. This new approach allows one to deal with recurrence systems with inhomogeneous parts in terms of indefinite nested sums that covers the pure holonomic and difference ring approaches as special cases. 
So far this approach has been non-trivially applied to obtain the first computer assisted proof~\cite{APS:05} of 
Stembridge's TSPP theorem~\cite{STEMBRIDGE1995227} and to provide the first proof of a non-trivial identity in~\cite{SZ:21} that is connected to irrationality proofs of zeta-values. In QCD-calculations this
new approach has been explored further to evaluate, e.g., bubble topologies~\cite{Behring:2013dga}.
\section{Symbolic integration}\label{Sec:Integration}

\vspace*{1mm}
\noindent
In the following we will present some of the most relevant tools of symbolic integration that have 
been used (at least in parts) in particular for multi--loop calculations in the case of a few number 
of external legs in elementary particle physics. Other tools suited for lower loop multi--leg 
calculations are described in part e.g. in \cite{Abreu:2022mfk}.
\subsection{The hyperlogarithm approach} 

\vspace*{1mm}
\noindent
If a Feynman diagram of the form~\eqref{Equ:GeneralForm} has no pole terms in~\eqref{Equ:EpExpansion} (i.e., $l=0$) or 
can be made finite by certain transformations splitting off its pole terms \cite{vonManteuffel:2014qoa}, it can be 
calculated under certain conditions by using the method of hyperlogarithms \cite{Brown:2008um}. 
Since here the denominator of the integral~\eqref{Equ:GeneralForm} is a multinomial in the Feynman parameters 
$x_i \in [0,1]$, one may seek a sequence of integrations, such that the denominator is always
a linear function in the integration variable. In this case the 
Feynman integral can be found as a linear combination of Kummer-Poincar\'{e} iterated integrals (also 
known as Goncharov iterated integrals), \cite{KUMMER:1840,POINCARE:1884,LAPPO:1953,CHEN:1971,Goncharov:1998}. The 
method has been first devised for massless scalar integrals in 
\cite{Brown:2008um}, for a corresponding code see \cite{Panzer:2014caa}, and it has been generalized
to massive diagrams~\cite{Ablinger:2014yaa},
dealing even with cases with no thorough multi-linearity, which is an extension to 
\cite{Brown:2008um,Panzer:2014caa}.
\hfill\\[-0.35cm]
\begin{figure}
	
	\vspace*{-0.3cm}
	
	\centering
	\includegraphics[width=3cm,height=1.4cm]{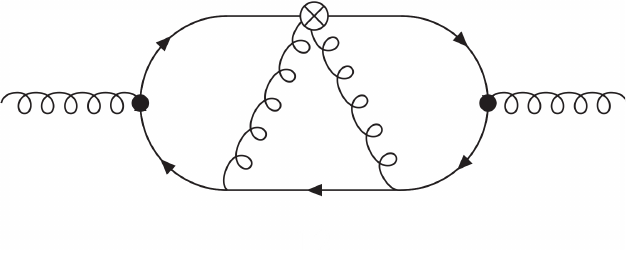}
	
	\vspace*{-0.5cm}
	
	\caption{\label{Fig:LadderDiagram}{A 3-loop ladder diagram with a central 
triangle~\cite{Ablinger:2015tua}.}}
	
	\vspace*{-0.2cm}
	
\end{figure}

\subsection{The multivariate Almkvist-Zeilberger approach}

\vspace*{1mm}
\noindent
Similar to the WZ summation approach its continuous version, the multivariate Almkvist-Zeilberger algorithm~\cite{AZ:06}, can compute 
a linear recurrence/difference equation
for a Feynman integral of the form~\eqref{Equ:GeneralForm}. 
Likewise, if the Feynman integral depends on a continuous parameter $x$ and the integrand is hyperexponential in $x$, one can search for a linear differential equation of the form
$$\sum_{i=0}^{\lambda}a_i(x,\ep)D_x^iF(x,\ep) =r(x,\ep).$$
A refined and improved method for the input class of Feynman integrals has 
been developed~\cite{Ablinger:12,Ablinger:2015tua,Ablinger:21}
which can hunt efficiently for homogeneous recurrences or differential equations. 
E.g., a recurrence in $n$ of order 5 can be calculated in about 8 hours for the master integral
$$\int_0^1\int_0^1\int_0^1\int_0^{1-u}\tfrac{
	(x+y-1)^N
	x^{\varepsilon/2} (1-x)^{\varepsilon/2} 
	y^{\varepsilon/2} (1-y)^{\varepsilon/2} 
	(1-u-v)^{N}\big(1-u \tfrac{x}{x-1}-v \tfrac{y}{y-1}\big)^{-1+3/2 \varepsilon} }{u^{1+\varepsilon/2} v^{1+\varepsilon/2}}
dx\,dy\,du\,dv$$
that arose in the context to tackle the highly non-trivial 3-loop Feynman diagram given in 
Fig.~\ref{Fig:LadderDiagram}; see~\cite{Ablinger:2015tua}. Then using the the linear difference equation solver 
of~\texttt{Sigma} one can compute the first coefficients of the $\ep$-expansion~\eqref{Equ:EpExpansion} in terms of harmonic sums and generalized harmonic sums. More generally, one can utilize the algorithms 
from~\cite{Blumlein:2010zv,Ablinger:21} to solve linear difference and differential equations in terms iterated sums and 
integrals; for 
further details see Section~\ref{Sec:LinearSolver}.

\subsection{The differential field and holonomic approach}

\vspace*{1mm}
\noindent
Risch's algorithm~\cite{Risch:69} for indefinite integration (for details see~\cite{Bron:97}) allows as input an integrand from the class of elementary functions (they are recursively built by compositions of
algebraic, exponential or logarithmic functions and the standard operations $+,-,\times,/$) and one can decide, 
if the indefinite integral defined over the input function can be written again in terms of elementary 
functions. Inspired by this result many further extensions have been derived. In particular, 
with~\cite{Singer:85,Bron:97} it is possible to deal with special classes of Liouvillian integrands 
(recursively built by indefinite integrals and hyperexponentials). In this regard, e.g., the package 
\texttt{Integrator}~\cite{Raab:12} enables one to treat not only indefinite integration problems, but also to 
compute difference/differential equations if the integrand depends on a discrete/continuous parameter. These 
tools have been exploited, e.g., to study root-valued integrals in~\cite{Ablinger:2014bra} that arise within 
massive 3-loop Feynman integral calculations.

In particular, the holonomic system approach~\cite{Zeilberger:90a,Chyzak:00,Koutschan:13} can be applied not only to multi-sums, but also to multi-integrals of the form~\eqref{Equ:GeneralForm} to determine a linear recurrence in a discrete parameter $n$ or a linear differential equation in a continuous parameter $x$. Analogously to the sum case, one can compute stepwise systems of linear differential/difference equations working from inside to outside of the multi-integral and ending up at a scalar equation of the free parameter $n$ or $x$. 
First examples have been elaborated in~\cite{Koutschan:13,Koutschan2021} using the package
\texttt{HolonomicFunctions} that illustrate further possibilities in QCD-calculations. 
\section{The method of arbitrarily large moments}
\label{Sec:largemoment}

\vspace*{1mm}
\noindent
One is often interested in the calculation of a certain number of moments in the 
Mellin variable, say $n=0,1,2,\dots,\mu$, to predict extra properties of physical quantities in terms of Feynman integrals.
Standard procedures, like {\tt Mincer} \cite{Mincer:91} or {\tt MATAD} \cite{Steinhauser:2000ry}, 
allow the calculation of a comparable small number of Mellin 
moments, e.g., $\mu=20$. Recently, a new method has been worked out in Ref.~\cite{Blumlein:2017dxp} 
and implemented within the package~\texttt{SolveCoupledSystem}~\cite{Ablinger:2016wlr,
Blumlein:2019oas,Blumlein:2019hfc} to compute thousands of 
such moments. 

In general, this new method assumes that we are given a coupled system~\eqref{eq:DEQ1} with~\eqref{Equ:GenerateFu} where already $\mu$ moments for the inhomogeneous part in~\eqref{eq:DEQ1} are computed (by applying this method recursively).
Then given such an input, one follows the calculation steps in Fig.~\ref{fig:uncouplingtactic} but instead of solving the recurrence in step~(3), one uses the recurrence together with a small number of initial values of $F_1(n)$ (bounded by the order of the recurrence) to compute in linear time the moments $F_1(0),\dots,F_1(\mu)$, and finally the corresponding moments for $F_2(n),\dots,F_{\lambda}(n)$. 
If the $F_i(n)$ depend also on $\ep$, one can calculate the moments of the coefficients of the $\ep$-expansions 
by exploiting refined ideas from~\cite{Blumlein:2010zv}.

More generally, using IBP methods~\cite{Chetyrkin:1981qh,Laporta:2001dd}, we suppose that a physical 
expression $\bar{f}(x)$ is given in terms of master integrals that are described in terms of recursively defined coupled systems of differential equations. Then using the large moment method iteratively one can calculate for a very large $\mu$ the moments of the master integrals. Assembling all the building blocks in the physical expression $\bar{f}(x)$, one finally derives at the coefficients
$F(0),\dots,F(\mu)$ of its power series~\eqref{Equ:OtherRepPS} in terms of rational numbers (if $\zeta$-values and other constants arise linearly, they are separated accordingly). 

\begin{figure}
	\footnotesize
	$$\xymatrix@C0.cm@R=0.8cm
	{
		&\txt{coupled systems}\ar[d]&\\
		&\txt{large no.~of moments}\ar[dl]\ar[d]^{\hspace*{-0.4cm} \txt{\hspace*{-0.5cm}\footnotesize \txt{\tt\scriptsize guessing}}}&\\
		\txt{numerics}&\txt{recurrence}
		\ar[ddl]^{\hspace*{-0.9cm}\hspace*{0.3cm}\txt{\tt\scriptsize available\\ \hspace*{-0.2cm}\tt\scriptsize solvers}}\ar[ddr]^(.68){\hspace*{-0.9cm}\txt{\hspace*{-0.1cm}\tt\scriptsize new\\ \hspace*{0.4cm}\tt\scriptsize solvers}}\\
		&&\\
		\txt{indefinite nested\\ 
			sums over hyper-\\geometric products}\hspace*{-1cm}&&\hspace*{-1.5cm}\txt{indefinite nested\\
			sums over ${}_pF_q$s\\ 
			(e.g., elliptic integrals)}\\
	}$$
	\normalsize
	\caption{\label{fig:largemoment}The large moment engine.}
\end{figure}
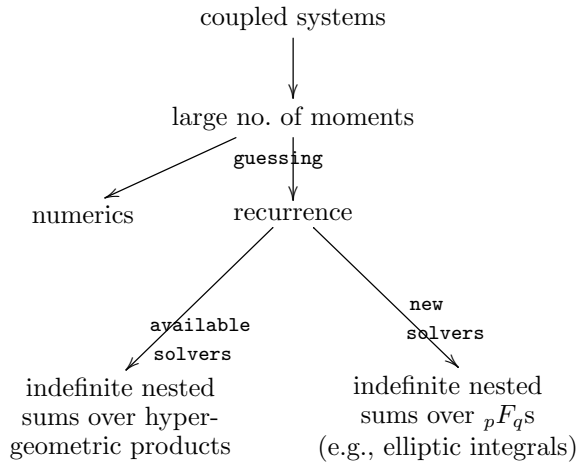

While in traditional solving methods very complicated function spaces might arise in intermediate steps, the large moment method deals simply with rational numbers and one can represent physical quantities with such sequences without entering any structural challenges.

Given a large number of moments, one may follow various strategies illustrated in Fig.~\ref{fig:largemoment}.
First, one can try to obtain interpolation expressions, e.g., by using orthogonal 
polynomials~\cite{Furmanski:1981ja} that provide numerical data of sufficient high precision being 
relevant for the experiments at the LHC and other future colliders.

Second, one can apply the guessing methods from Section~\ref{Sec:GuessREDE} in order to produce linear recurrences with minimal order for the physical quantities. In short, analyzing this quantity amounts precisely to the exploration of the computed recurrence.\\
Next, one can try to solve the recurrences in terms of special functions by using the tools form Section~\ref{Sec:LinearSolver}.
This strategy is particularly successful if the final result (but not necessarily the intermediate results) can be given in terms of indefinite nested sums over hypergeometric products. As demonstrated 
in~\cite{Blumlein:2009tj}, we could calculate from about $\mu=5000$ moments all the recurrences that determine 
the massless
unpolarized 3-loop anomalous dimensions and Wilson coefficients in deep-inelastic scattering
\cite{Moch:2004pa,Vogt:2004mw,Vermaseren:2005qc} by solving the recurrences. 
Similarly, we could calculate, e.g., the 3-loop splitting functions~\cite{Ablinger:2017tan},
the massive 2- and 3-loop form factor~\cite{Ablinger:2018yae,Blumlein:2019oas}, 
the anomalous dimensions from off shell operator 
matrix elements~\cite{Blumlein:2021enk,Behring:2019tus,Blumlein:2021ryt},
the polarized transition matrix element $A_{gq}(N)$~\cite{Behring:2021asx} and others, the logarithmic 
contributions to the 
polarized $O(\alpha_s^3)$ asymptotic massive Wilson coefficients~\cite{Blumlein:2021xlc},
and the two-loop massless 
off-shell QCD operator matrix elements~\cite{Blumlein:2022ndg}. \\
We remark that the found recurrences may also contribute substantially in the case that one fails to find closed 
form 
solutions. For instance, one may extract the asymptotic behavior of the physical quantities by using methods described in~\cite{WZ:85,Kauers:11}.

\section{Special functions and their algorithms}\label{Sec:SpecialFunctions}

\vspace*{1mm}
\noindent
The representation of the results of calculations in QCD and QED are 
characterized by special constants and functions. The former ones appear in zero scale calculations
and as boundary conditions in single and more scale problems. Since in particular in QCD and QED 
the Mellin transform (see~\eqref{Equ:OtherRepM} where in the following $x$ is replaced by $z$)
relates nested sums at the one hand to nested integrals at the other hand, and vice versa, two 
principle classes
of special single scale functions emerge: indefinitely nested sums over hypergeometric products and 
iterated integrals over certain alphabets of letters. Both in the limit $n \rightarrow \infty$ of 
the sums and at $z=1$ for the iterated integrals special numbers are obtained. Examples on 
different classes of functions are given in Tab.~\ref{fig:SpecFu}.
All these function spaces obey (quasi) shuffle relations, cf.~\cite{Hoffman,Blumlein:2003gb}, 
implying algebraic relations, which allow to reduce to the respective algebraic bases 
\cite{Blumlein:2003gb,AS:18}.
\begin{table}
	\centering
	\footnotesize
	\begin{tabular}{lll}
		{\fs \bf Nested sums}\hspace*{1.5cm} & {\fs \bf Nested integrals}\hspace*{2.5cm} & {\fs \bf Special 
			numbers}\\
		\hline\hline
		Harmonic Sums & Harmonic Polylogarithms & multiple zeta values \\[-0.1cm]
		\scriptsize{$ \displaystyle\sum_{k=1}^n\frac{1}{k} \sum_{l=1}^k \frac{(-1)^l}{l^3}$} & 
		{\scriptsize$ \displaystyle \int_0^z \frac{dy}{y} \int_0^y \frac{dx}{1+x}$} &
		{\scriptsize$ \displaystyle \int_0^1 dx \frac{\Li_3(x)}{1+x} = -2 \Li_4(1/2) + ...$}\\
		\hline
		gen. Harmonic Sums & gen. Harmonic Polylogarithms & gen. multiple zeta values \\[-0.1cm]
		{\scriptsize$ \displaystyle \sum_{k=1}^n \frac{(1/2)^k}{k} \sum_{l=1}^k \frac{(-1)^l}{l^3}$} 
		& 
		{\scriptsize$ \displaystyle \int_0^z \frac{dy}{y} \int_0^y \frac{dx}{x-3}$} &
		{$ \displaystyle \int_0^1 dx \frac{\ln(x+2)}{x-3/2}  = \Li_2(1/3)+ ...$}
		\\
		\hline
		Cycl. Harmonic Sums & Cycl. Harmonic Polylogarithms & cycl. multiple zeta values \\[-0.1cm]
		{\scriptsize$ \displaystyle \sum_{k=1}^n \frac{1}{(2k+1)} \sum_{l=1}^k \frac{(-1)^l}{l^3}$} & 
		{\scriptsize$ \displaystyle \int_0^z \frac{dy}{1+y^2} \int_0^y \frac{dx}{1 -x + x^2}$} &
		{\scriptsize$ \displaystyle {\bf C} = \sum_{k=0}^\infty \frac{(-1)^k}{(2k+1)^2} 
			$}
		\\
		\hline
		Binomial  Sums & root-valued iterated integrals & associated numbers \\[-0.1cm]
		{\scriptsize$ \displaystyle \sum_{k=1}^n \frac{1}{k^2} \binom{2k}{k} (-1)^k$} & 
		{\scriptsize$ \displaystyle \int_0^z \frac{dy}{y} \int_0^y \frac{dx}{x\sqrt{1+x}}$} &
		{\scriptsize$ \displaystyle {\rm H}_{8,w_3} = 2 {\rm arccot}(\sqrt{7})^2 $}\\
		\hline
		&iterated integrals on $_2F_1$'s & associated numbers \\[-0.cm]
		&
		\hspace*{-0.4cm}{\scriptsize$ 
			\displaystyle \int_0^z
			\frac{\ln(x)}{1+x} 
			\pFq{2}{1}{{\frac{4}{3}},\frac{5}{3}}{2}{\frac{x^2(x^2-9)^2}{(x^2+3)^3}}dx
			$} &
		{\scriptsize$ \displaystyle 
			\displaystyle \int_0^1  
			\pFq{2}{1}{{\frac{4}{3}},\frac{5}{3}}{2}{\frac{x^2(x^2-9)^2}{(x^2+3)^3}}dx
			$}\\
		\hline
	\end{tabular}
	\normalsize
	
	\vspace*{-0.25cm}
	
	\caption{\label{fig:SpecFu}{Special functions and numbers.}}
	
	\vspace*{-0.65cm}
	
\end{table}

Historically, most of the Feynman diagram calculations in the time before 1998 were performed using
$z$--space representations leading to classical polylogarithms and Nielsen integrals 
\cite{NIELSEN:1909,Kolbig:1969zza,Kolbig:1983qt,LEWIN1,LEWIN2,Devoto:1983tc},
partly with involved arguments. 

A systematic description in terms of harmonic sums started in 1998
with \cite{Blumlein:1998if,Vermaseren:99}. They are defined by 
\begin{eqnarray}
S_{b,\vec{a}}(n) = \sum_{k=1}^n \frac{({\rm sign}(b))^k}{k^{|b|}} S_{\vec{a}}(k),~~S_\emptyset 
=1,~~
b, a_i \in \mathbb{Z} \backslash \{0\}, 
n \in \mathbb{N} \backslash \{0\}.
\end{eqnarray}
Related to that, the iterative integrals are the harmonic polylogarithms,
\begin{eqnarray}
\HA_{b,\vec{a}}(z) = \int_0^z dy f_b(y) \HA_{\vec{a}}(y),~
\HA_\emptyset =1,~~b, a_i \in \{0,1,-1\},
\end{eqnarray}
with the alphabet
\begin{eqnarray}
\mathfrak{A} = \left\{
f_0(z) = \frac{1}{z}, 
f_1(z) = \frac{1}{1-z}, 
f_{-1}(z) = \frac{1}{1+z} \right\},
\end{eqnarray}
\cite{Remiddi:1999ew}. A special defintion is required for the case $\HA_{\vec{a}}(z), \forall a_i = 
0, i = 1...n$, which has no integral representation, but is defined as $\ln^n(z)/n!$, for 
completeness. In the case of infinite sums we also allow for the 
symbol $\sigma_\infty :=
\sum_{k=1}^\infty (1/k)$, which is not a number, but simplifies various algebraic relations and is 
therefore useful. 

The special numbers are multiple zeta values in both cases. Their 
representations at high 
order can be found in \cite{Blumlein:2009cf}. 

At the next level, generalized harmonic sums \cite{Moch:02,Ablinger:2013cf} contribute, e.g.\ in the 
case of the pure singlet 3--loop massive Wilson coefficients in the asymptotic region 
\cite{Ablinger:2014nga}. These quantities are given by
\begin{eqnarray}
S_{b,\vec{a}}(\{c,\vec{d}\};n) &=& \sum_{k=1}^n \frac{c^k}{k^{b}} 
S_{\vec{a}}(\{\vec{d}\};k),~~S_\emptyset  
= 1,
\nonumber\\ &&
b, a_i \in \mathbb{N} \backslash \{0\}, c, d_i \in \mathbb{C} \backslash \{0\},
n \in \mathbb{N} \backslash \{0\}.
\end{eqnarray}
The corresponding iterated integrals are also called Kummer--Poincar\'e iterated integrals
\cite{KUMMER:1840,POINCARE:1884,LAPPO:1953,CHEN:1971,Goncharov:1998}
and are given by
\begin{eqnarray}
\HA_{b,\vec{a}}(z) = \int_0^z dy f_b(y) \HA_{\vec{a}}(y),~
\HA_\emptyset =1,
\end{eqnarray}
with the alphabet
\begin{eqnarray}
\mathfrak{A} = \left\{f_{c_i}(z) = \frac{1}{z - c_i},~~c_i \in \mathbb{C} \right\}.
\end{eqnarray}

Further, cyclotomic harmonic sums and polylogarithms~\cite{Ablinger:2011te} contribute.
The letters of the alphabet forming the iterated integrals are those of the harmonic polylogarithms
extended  with letters of the type
\begin{eqnarray}
f_{k,a}^{\rm cycl.}(z) = \frac{z^a}{P_k(z)},~~k \geq 3,
\end{eqnarray}
with $k$ labeling the cyclotomic polynomials and $a \in [0,\varphi(k)]$, and $\varphi(k)$ is Euler's 
totient function. The associated cyclotomic harmonic sums iterate monomials of the type
\begin{eqnarray}
\frac{s^k}{(a k + b)^c},~~a,c \in \mathbb{N}_+, b \in \mathbb{N}, s \in \mathbb{C} \backslash \{0\}.
\end{eqnarray}

Finite binomial sums \cite{Ablinger:2014bra} contribute for a series of 
topologies in the massive OMEs $A_{gg}^{(3)}$ \cite{Ablinger:2014uka}
and $A_{Qg}^{(3)}$ \cite{Ablinger:2015tua}.
The corresponding sums are generalized sums with an additional factor of $\binom{2k}{k}$ in the 
numerator or denominator. The associated iterated integrals, obtained by a Mellin inversion,
are formed out of letters containing square root valued structures, as e.g.\ shown in 
Tab.~\ref{fig:SpecFu}. Another example is
\begin{multline*}
\sum _{i=1}^n \frac{1}{\binom{2 i}{i}i^3}=\sum _{i=1}^{\infty} \frac{1}{\binom{2 i}{i}i^3}\\
+\frac1{4^{n}} \int_0^1 z^n \Big(\tfrac{3\ln(z)^2-12\ln (2) \ln(z)+12\ln (2)^2 -\pi ^2}{6 (-4+z)}+\tfrac{\int_0^z\frac{1}{\tau_1}\int_0^{\tau_1}\frac{\sqrt{1-\tau_2}-1}{\tau_2}d\tau_2d\tau_1-2
		\int_0^z\frac{\sqrt{1-\tau}-1}{\tau}d\tau}{-4+z}\Big)\, dz.
\end{multline*}
\noindent In particular, iterative application of integration by parts yield the asymptotic expansion
\begin{equation}\label{Equ:Prop:BinomExpand}
\sum _{i=1}^n \frac{1}{\binom{2 i}{i}i^3}\sim 2^{-2 n} \sqrt{n} \sqrt{\pi } \Big(-\tfrac{34924547}{884736 n^7}+\tfrac{91999}{9216 n^6}-\tfrac{10537}{3456 n^5}+\tfrac{77}{72 n^4}-\tfrac{1}{3 n^3}+O\big(\tfrac{1}{n^8}\big)\Big)+\sum _{i=1}^{\infty} \tfrac{1}{\binom{2 i}{i}i^3}.
\end{equation}
\noindent 
Such expansions are extremely useful for limit calculations and for analyzing the expression behavior 
for large values of $n$. Moreover, the sum and integral representations equipped with their shuffle and 
quasi-shuffle algebras~\cite{Hoffman,Blumlein:2003gb,Blumlein:2009ta,Blumlein:2009fz} give rise to algebraic 
relations 
of infinite sums. In particular, attaching special constants to sums that cannot be simplified further,
one can discover evaluations such as 
$$\sum_{i=1}^{\infty}\frac{2^iS_1(i)}{i\binom{2i}{i}}=2\,C-\frac{\pi\,\log(2)}{2}+\frac{3}{4}\zeta(2)$$
\noindent where $C=\sum_{i=1}^{\infty}\frac{(-1)^i}{(2i+1)^2}$ denotes the Catalan constant; 
for further details see~\cite{Ablinger:19,Ablinger:20}.
For general classes, like nested binomial sums, more flexible methods were developed recently 
to map between $n$- 
and $z$-space, cf.~\cite{Ablinger:2018brx}: 
given a recurrence of $F(n)$ (resp. a differential equation of $f(z)$), compute a differential equation for $f(z)$ 
(resp.\ a recurrence for $F(n)$). In particular, using the introduced solvers from Section~\ref{Sec:OrdinaryEquation}, 
one can check, if the Mellin transform (resp.\ inverse Mellin transform) can be given in terms of indefinite nested 
sums (resp.\ integrals). Infinite (inverse) binomial sums have been also studied in 
\cite{Davydychev:2003mv,Weinzierl:2004bn}. For the simpler cases efficient rewrite rules have been 
developed to switch between the sum and integral representations via the (inverse) Mellin transform. 

In more general cases, in particular in two--scale problems, the so called $G$--functions appear, 
which are iterated integrals over larger alphabets, partly with root--valued letters. 
They are given by
\begin{eqnarray}
\label{eq:GF}
G\left(f_a(x),f_{b_1}(x),...,f_{b_n}(x)\right) = \int_0^x dy 
f_a(y) G\left(f_{b_1}(y),...,f_{b_n}(y)\right).
\end{eqnarray}
Actually $f_c(x)$ even denotes in general a differentiable function, up to regularizations in special cases. 

At higher and higher orders in perturbation theory, new building blocks arise that cannot be 
represented in terms of indefinite nested sums or iterated integrals. In particular, one ends up at 
linear difference/differential equations, that cannot be solved completely in terms of 
d'Alembertian/Liouvillian solutions. For this reason, the class of iterative non--iterative 
integrals have been introduced in 2016 \cite{Blumlein:2016}.

Probably the first case in which complete elliptic integrals emerged in a quantum field 
theoretic calculation has been the fourth order spectral functions for the electron propagator by 
Sabry 1962 \cite{SABRY}. For complete physical processes more recently elliptic integrals were 
needed. This is the case in massive three--loop calculations for the QCD corrections of the $\rho$ 
parameter
\cite{Ablinger:2017bjx,Blumlein:2018aeq} 2017\footnote{The same differential equations rule the
non first-order factorizing cases in the calculation of the massive three--loop operator matrix 
element $A_{Qg}$ in the single mass case \cite{Blumlein:2017wxd}, found together with the 
calculation \cite{Blumlein:2018aeq}.} and in massless three--three loop calculations 2018 
and later \cite{Mistlberger:2018etf,Duhr:2020seh}. 
There is a series of well-known examples of individual integrals of a certain structure in the 
literature as the sun-rise integral, 
cf.~e.g.~\cite{Broadhurst:1993mw,Bloch:2013tra,Adams:2015ydq} and
the kite-integral \cite{SABRY,Remiddi:2016gno,Adams:2016xah}; for a collection of recent surveys see 
Ref.~\cite{Blumlein:2019tmi}.\footnote{Very naturally, as now the technical aspects on complete 
elliptic integrals are very well known in particle physics, many applications find these 
contributions, cf. e.g. \cite{Henn:2020lye,Bern:2021dqo,Bargiela:2021wuy}, 
the reason being the occurrence of 
the corresponding Heun and $_2F_1$-type differential equations, cf. e.g.~\cite{Ablinger:2017bjx}.}
In general, these classes of integrals form iterative 
non--iterative integrals, cf.~\cite{Ablinger:2017bjx}. Beyond this level one has Abel--integrals 
\cite{Neumann:1884} and Calabi--Yau structures, cf.~\cite{Brown:2010bw,Bonisch:2021yfw,Kreimer:2022fxm}. Even more
involved structures will occur at higher topologies. In Mellin space they have the common characteristics of 
difference equations with rational coefficients which are not factorizing at first order. Any of the corresponding 
solutions also needs efficient numerical representations, as 
e.g.~\cite{Gehrmann:2001pz,Vollinga:2004sn,Ablinger:2018sat}, for phenomenological and experimental
applications. 
This also applies to Mellin space representations for $n \in \mathbb{C}$, 
\cite{Blumlein:2000hw,Blumlein:2005jg,Kotikov:2005gr,Blumlein:2009ta,Blumlein:2009fz}.
Feynman integrals will imply a multitude of new function spaces in the future.
\section{Calculations in Quantum Field Theory}
\label{sec:CQFT}

\vspace*{1mm}
\noindent
Our major topic concerns analytic Feynman diagram calculations. As has been shown, this is deeply 
rooted in solving large systems of differential or difference equations. The single scale cases are 
mathematically widely understood and one may project to the zero scale case, i.e.\ to special numbers.
However, just by this one will probably not be able to find all relations 
between these special numbers by
using, e.g., the techniques described in Section~\ref{Sec:IntegerRel}, 
beginning at a certain level of complexity, which requires further advanced methods in these cases.
On the other side, as experience shows, certain two--scale problems can still be solved 
analytically, 
as we will discuss in Section~\ref{Sec:CQFT3}. But already starting at that level, one has to deal 
with  
partial differential and difference equations, on which is much less known, cf.~\cite{Lie:1891,Schwarz:2008}. 
In the single scale case, going to higher and higher orders, one will face non--first order factorizing 
differential and difference equations of higher and higher order 
\cite{Tricomi:1948,CS:2017,Neumann:1884}\footnote{For more literature on elliptic integrals and modular forms 
see Refs.~\cite{Blumlein:2018cms,Ablinger:2017bjx,Blumlein:2019tmi}.}, for which only 
the properties of very few concrete classes have been studied so far and a very wide field of future 
mathematical investigation is opening up. 

For more scales, one probably will have to rely on using numerical precision methods in the first 
place, because of the wide variety of structures \cite{Heinrich:2020ybq}. Computational Quantum Field 
Theory (QFT) is urged to invest 
much more 
efforts to obtain fast and highly reliable methods in this direction to be able to cope with the 
challenges in future precision measurements. Developments of this kind may take quite a long time and need
intense collaboration with experts in the field of numerical mathematics.

\subsection{Zero Scale Calculations}
\label{Sec:CQFT0}

\vspace*{1mm}
\noindent
Zero scale quantities in QFTs, as QED and QCD, are characterized by color factors, rational coefficients 
and special numbers like multiple zeta values \cite{Blumlein:2009cf}. Examples are fixed moments for
massive three--loop OMEs \cite{Bierenbaum:2009mv} and massless four--loop anomalous dimensions
\cite{Moch:2021qrk}. Particularly for massive problems more special numbers contribute, as those 
related to generalized harmonic sums 
\cite{Ablinger:2013cf},
cyclotomic harmonic sums \cite{Ablinger:2011te}, binomial sums \cite{Ablinger:2014bra}, and those 
related to elliptic integrals \cite{Laporta:2017okg}, see also Tab.~\ref{fig:SpecFu}. More and more different sets 
will 
emerge including even higher topologies. One also may calculate moments of single scale quantities, which 
depend on an integer parameter $n$, by obtaining sequences of rational numbers. These numbers incorporate thus
an essential part of the more involved single scale dependence for general values of $n$. It is sometimes of
advantage to first work with these moments, despite the fact that the general $n$ relation is determined
by a difference equation, which does not factorize at first order. This is often the case for master integrals 
in the massive case. However, the corresponding recurrences for anomalous dimensions are factorizing at first 
order. One inserts first the master integrals for fixed moments and then determines the difference equation for 
the anomalous dimension, see \cite{Ablinger:2017tan,Behring:2019tus}.
\subsection{Massless Single Scale Calculations}
\label{Sec:CQFT1}

\vspace*{1mm}
\noindent
These quantities are the anomalous dimensions, currently known to three--loop order 
\cite{Moch:2004pa,Vogt:2004mw,Moch:2014sna,Ablinger:2014nga,Ablinger:2017tan,Behring:2019tus,
Blumlein:2021enk,Blumlein:2021ryt}, the massless Wilson coefficients for deep--inelastic scattering 
\cite{Blumlein:2012bf}
up to $O(\alpha_s^3)$ \cite{Vermaseren:2005qc}, the Drell--Yan 
process and Higgs production to two--loop order 
\cite{Hamberg:1990np,Harlander:2002wh,Ravindran:2003um,Blumlein:2005im}. All these quantities 
can be expressed by harmonic sums \cite{Vermaseren:99,Blumlein:1998if} in Mellin space and by 
harmonic polylogarithms in $z$--space \cite{Remiddi:1999ew}. For Higgs production and the  
Drell--Yan process at three--loop order \cite{Mistlberger:2018etf,Duhr:2020seh} also elliptic 
integrals contribute. It is generally expected
that a further nesting in the Feynman diagram topologies leads to new mathematical structures also in 
the massless case in higher orders of the coupling constant.

For massless single scale calculations one may very efficiently apply the method of arbitrarily high 
moments \cite{Blumlein:2017dxp}, together with guessing to obtain the recurrences, which may either
be solved or reduced, by factoring of the first order factors, using {\tt Sigma}. This also applies 
to the case of massive single scale calculations, to which we turn now. 
\subsection{Massive Single Scale Calculations}
\label{Sec:CQFT2}

\vspace*{1mm}
\noindent
The method of massive OMEs \cite{Buza:1995ie} allowed to calculate single scale quantities, such as 
the heavy 
flavor Wilson coefficients to three--loop order in the asymptotic region, obtaining all logarithmic 
contributions \cite{Behring:2014eya,Blumlein:2021xlc} and also the constant term. The method has 
also been applied to problems in 
QED, cf.~\cite{Blumlein:2011mi,Ablinger:2020qvo,Blumlein:2021jdl}, see Section~\ref{Sec:CQFT21}.
In some cases even full results have been obtained at two--loop order 
\cite{Buza:1995ie,Blumlein:2016xcy,Blumlein:2019qze,Blumlein:2019zux,Blumlein:2011mi}, cf. 
Section~\ref{Sec:CQFT22}.
\subsubsection{Massive Single Scale Calculations: logarithmic and constant corrections}
\label{Sec:CQFT21}

\vspace*{1mm}
\noindent 
The asymptotic heavy flavor Wilson coefficients of deep--inelastic scattering contain single 
scale logarithmic and constant contributions. At two--loop order all contributions are known 
\cite{Buza:1995ie,Buza:1996xr,Buza:1996wv,Bierenbaum:2007qe,Bierenbaum:2008yu,Bierenbaum:2009zt}. 
At three--loop order all but the massive OME $A_{Qg}^{(3)}$ have been calculated 
analytically in complete form both in the unpolarized and polarized case. In Mellin $n$ space 
they can be expressed by harmonic sums for all $N_F$-terms \cite{Ablinger:2010ty,Blumlein:2012vq}, 
and for $A_{qq,Q}^{(3),\rm NS}$, $A_{qq,Q}^{(3),\rm PS}$, $A_{qg,Q}^{(3)}$ and $A_{gq,Q}^{(3)}$ 
\cite{Ablinger:2014vwa,Behring:2014eya,Ablinger:2014lka,Ablinger:2020snj,Blumlein:2021xlc}. 
Generalized harmonic sums 
contribute in the pure singlet case $A_{Qq}^{(3),\rm PS}$ \cite{Ablinger:2014nga,Ablinger:2019etw} and finite 
binomial sums for
$A_{gg,Q}^{(3)}$ \cite{Ablinger:2014uka}. Finally, $A_{Qg}^{(3)}$ receives also contributions by 
complete elliptic integrals \cite{Blumlein:2017wxd}.

Another case belonging to this class of integrals are the contributions to the massive form factor at 
three--loop order. The first order factorizing contributions can be expressed by harmonic and 
cyclotomic harmonic polylogarithms in the variable $x$
\begin{eqnarray}
\frac{q^2}{m^2} = - \frac{(1-x)^2}{x},
\end{eqnarray}
with $q^2$ the virtuality and $m$ the heavy quark mass, 
\cite{Ablinger:2018yae,Blumlein:2018tmz,Ablinger:2018zwz,Lee:2018rgs,Lee:2018nxa,Blumlein:2019oas}

The method of massive OMEs has also been applied for the calculation of the initial state 
radiation to the process $e^+e^- \rightarrow Z^*/\gamma^*$. These corrections are of importance
for planned future high luminosity measurements at the ILC, CLIC, and FCC\_ee. The results at 
$O(\alpha^2)$ \cite{Blumlein:2011mi} showed disagreement with an earlier direct calculation
\cite{Berends:1987ab}. The only way to find the correct results has been a complete diagrammatic 
calculation, without expansion in the small parameter $\rho = m_e^2/s$, with $m_e$ the electron 
mass and $s$ the cms energy. This has been performed in Ref.~\cite{Blumlein:2020jrf}. Furthermore,
we expanded in $\rho$ through different steps, controlled by high precision numerics, and confirmed 
the results of \cite{Blumlein:2011mi}. The fermionic integrals could be represented using iterated 
Kummer--elliptic integrals over larger alphabets. Numerical results were presented in 
\cite{Blumlein:2019srk,Blumlein:2019pqb}. The method of massive OMEs has then been extended to 
calculate the first three logarithmic series up to $O(\alpha^6 L^5)$, where $L = \ln(s/m_e^2)$ 
in \cite{Ablinger:2020qvo}. Here in Mellin space also generalized harmonic sums contribute.
Higher order corrections for the forward--backward asymmetry were calculated in 
\cite{Blumlein:2021jdl}, where also cyclotomic harmonic polylogarithms contribute to the 
radiators.\footnote{For a recent survey on the QED corrections see \cite{Blumlein:2022mrp}.} 
\subsubsection{Massive Single Scale Calculations: including also power corrections}
\label{Sec:CQFT22}

\vspace*{1mm}
\noindent
In some cases, massive two--loop problems can be integrated analytically in the whole kinematic region.
This applies to the flavor non--singlet contributions \cite{Buza:1995ie,Blumlein:2016xcy} and 
the pure singlet contributions \cite{Blumlein:2019qze,Blumlein:2019zux}. While in the non--singlet case
classical polylogarithms with root--valued arguments suffice for the representation, in the pure singlet case
iterated integrals over alphabets containing Kummer--elliptic letters are necessary. Part of them integrates to 
incomplete elliptic integrals, which do not destroy the iterated integral structure, 
unlike the case in the iterative non--iterative integrals \cite{Blumlein:2017rzm}. 
In establishing the contributing alphabet also rationalization of roots is performed as far as 
possible; for other investigations see also \cite{Besier:2019kco}. Examples for these letters are, 
cf.~\cite{Blumlein:2019qze}
\begin{eqnarray}
f_{w_{11}}(t) &=& \frac{t}{\sqrt{1-t^2} \sqrt{1-k^2 t^2}}
\\
f_{w_{12}}(t) &=& \frac{t}{\sqrt{1-t^2} \sqrt{1-k^2 t^2}(k^2(1-t^2(1-z^2))-z^2)},
\end{eqnarray}
with $k = \sqrt{z}/{\sqrt{1-(1-z)\beta^2}}, \beta = \sqrt{1 - 4 m^2 z/(Q^2(1-z)}$ and $z$ is the 
momentum fraction variable. Depth-three iterated integrals over the contributing integrals emerge.

The fact that one finds analytic integral representations in these cases is related to the tree--like structure
of the contributing diagrams. At higher orders or for other processes, correspondingly, one has to perform 
corresponding expansions in $m^2/Q^2$, to successively obtain analytic results, improving the logarithmic and 
constant orders obtained in the region $Q^2 \gg m^2$. The possibility to analytically calculate the pure--singlet 
corrections, conjectured  
by van Neerven and J.B. around 2000, turned out to be correct, however, the necessary technologies for
this became only available with \cite{Blumlein:2019qze} later and Nielsen integrals with whatsoever 
complicated argument are not sufficient to represent the final result.

\vspace*{1mm}
\noindent

\subsection{Massive Double Scale Calculations}
\label{Sec:CQFT3}

\vspace*{1mm}
\noindent
In the case of deep--inelastic scattering from the level of three loop onward, diagrams contribute,
which contain charm and bottom quarks. This leads to a double scale problem, as similarly also in the 
case of the massive from factor and for other processes. In the following we will consider the case of 
deep--inelastic scattering. In all cases but the massive OMEs $A_{Qg}^{(3)}$ and its polarized 
counterpart $\Delta A_{Qg}^{(3)}$, complete analytic solutions are possible either in Mellin $n$ or 
momentum fraction $z$--space. The different OMEs have been calculated in 
Refs.~\cite{Ablinger:2017err,Ablinger:2017xml,Ablinger:2018brx,Blumlein:2018jfm,
Ablinger:2019gpu,Ablinger:2020snj}. They can be expressed by $G$--functions, cf. (\ref{eq:GF}).
One example is \cite{Ablinger:2018brx}
\begin{align}
G_{37} &= 
G\left(\left\{\frac{1}{1-x+\eta x},\sqrt{1-x} \sqrt{1-x+\eta x}\right\},z\right) 
\nonumber\\ &
= -\frac{\eta^2}{(1-\eta )^{5/2}} \Biggl\{
\frac{1}{16} \ln\left(2-\eta+2 \sqrt{1-\eta}\right)
+\frac{1}{4} \arcsin^2\left(\frac{1}{\sqrt{\eta }}\right)
\nonumber\\ &
+\frac{i}{2} \left[-\ln \left(1-\sqrt{1-\eta }\right)+\ln (\eta )-\ln (2)\right] 
\arcsin\left(\frac{1}{\sqrt{\eta }}\right)
\nonumber \\ & 
+\frac{1}{4} \text{Li}_2\left(\frac{\eta +2 \sqrt{1-\eta }-2}{\eta }\right)
+\frac{i}{2} \arcsin\left(\frac{\sqrt{1-z+\eta z}}{\sqrt{\eta }}\right) \ln \left(1+\chi\right)
\nonumber \\ & 
-\frac{1}{8} \ln \left(\sqrt{1-\eta } \sqrt{1-z}+\sqrt{1-z+\eta z}\right)
-\frac{1}{4} \arcsin^2\left(\frac{\sqrt{1-z+\eta z}}{\sqrt{\eta }}\right)
\nonumber \\ & 
+\frac{1}{4} \ln\big(1-z+\eta z\big) \biggl[
-\ln\left(1-\eta +\sqrt{1-\eta}\right)
+\frac{\ln(\eta)}{2}
+\frac{1}{2} \ln(1-\eta)
-\frac{i \pi }{2}
\biggr]
\nonumber \\ & 
-\frac{1}{4} \text{Li}_2\left(-\chi\right)
\Biggr\}
+\frac{3 \eta-2}{8 (1-\eta )^2}
-\frac{(2-\eta )}{4 (1-\eta)^2} \ln\big(1-z+\eta z\big)
\nonumber \\ & 
+\sqrt{1-z} \sqrt{1-z+\eta z} \frac{(2-3 \eta +2 \eta z-2 z)}{8 (1-\eta)^2} \, ,
\end{align}
where $\eta = m_1^2/m_2^2$ and $\chi = (1/\eta)(\sqrt{1-\eta}\sqrt{1-z} - \sqrt{1-z+\eta z})^2$.
In the case of the pure singlet two--mass contributions \cite{Ablinger:2017xml,Ablinger:2019gpu}
we work in $z$--space by using Mellin--Barnes integrals \cite{MELLIN,BARNES1,BARNES2}. One also obtains 
$G$--functions and in part integrals over them, with a different support than usual, expressed by
Heaviside functions. These problems cannot be solved in Mellin $n$ space.
\subsection{Classical Gravity}
\label{Sec:CQFT4}

\vspace*{1mm}
\noindent
The classical kinematics of massive astrophysical objects, such as black holes and neutron stars, can 
be calculated by using methods of effective field theory developed for Quantum Field Theory. 
Concepts 
like the path integral \cite{PATH} and Feynman diagrams are also applicable at the classical level.
This is an enormous bonus to the field of general relativity and classical gravity, since 
very advanced computation technologies, starting from Feynman diagram generation 
\cite{Nogueira:1991ex}, effective 
performance of Lorentz algebra \cite{Vermaseren:2000nd,Tentyukov:2007mu}, integration-by-parts 
reduction 
\cite{Marquard:2021spf},
and the calculation of master integrals already exist. One expands Einstein--Hilbert gravity in terms 
of auxiliary fields \cite{Kol:2007bc}. 
Furthermore, one performs the classical limit using the method of 
expansion by regions, cf.~\cite{Jantzen:2011nz,Blumlein:2020pyo}, where only the potential and 
radiation modes are 
contributing. These methods can be applied for the inspiraling process of the massive objects
\cite{Foffa:2019hrb,Blumlein:2019zku,Blumlein:2020pog,Blumlein:2021txj,Blumlein:2021txe,Bini:2020hmy}, 
as well 
as for their scattering process, cf.~\cite{Bern:2021yeh,Dlapa:2021vgp}. The main challenge 
is here to deal with the ever growing effective vertex structures and the integration by parts 
reduction, which can be performed by packages like {\tt Crusher}, cf.~\cite{Marquard:2021spf}.
The bound state kinematics is described by the post--Newtonian (PN) approach, having reached now 6PN
order \cite{Blumlein:2020znm,Bini:2020hmy, Blumlein:2021txj} and the scattering process by the 
post--Minkowskian approach, now available at $O(G_N^4)$, where $G_N$ denotes Newton's constant 
\cite{Bern:2021yeh,Dlapa:2021vgp}. After expanding the potential contributions of the 
post--Minkowskain results, a part of the PN results is re-obtained, which has been shown to 6PN in 
\cite{Blumlein:2020znm,Blumlein:2021txj}. This applies to the potential contributions.
In principle, the method of guessing could be used obtaining
post--Minkowskian results, again from potential contributions, \cite{Blumlein:2019bqq}. While 
agreement has been reached for the 4PN level between various approaches, the level of 5PN is still 
under discussion because of the non--potential contributions. For their description the different 
methods proposed in the literature do not lead to consistent results as of yet, requiring both more 
clear theoretical foundations and also more work to obtain the final result for bound state problems.
\section{Conclusion}
\label{Sec:Conclusion}

\begin{figure}
\centering
\footnotesize
$$\hspace*{-0.15cm}\xymatrix@!R=0.8cm@C0.1cm{
&&&\ar[dll]\ar[ddddll]\ar@/^1.5pc/[dddlll]\ar@/^-2.5pc/[ddlll]\boxed{\txt{system\\solving}}\ar[drr]\ar[ddddrr]\ar@/^-1.5pc/[dddrrr]\ar@/^2.5pc/[ddrrr]\ar@/^1.5pc/[drrr]&&&\\
&\ar@/^-0.8pc/[dl]\ar@/^0.9pc/[ddl]\ar[ddd]\boxed{\txt{symbolic\\integration}}&&&&\boxed{\txt{symbolic\\summation}}\ar[dr]\ar[ddr]\ar[ddd]&\hspace*{0.65cm}\boxed{\txt{computing\\moments}}\ar[d]\\
\boxed{\txt{differential\\equ.\ 
guessing}}\ar@/^0.8pc/[ur]\ar[ddr]&&&&&&\boxed{\txt{recurrence\\ guessing}}\ar[ul]\ar[ddl]\\
\boxed{\txt{differential\\ equ.\ 
solving}}\ar[dr]&&&*+[F=:<3pt>]{\txt{physical\\problems}}\ar@{.>}[uuu]\ar@{.>}[dd]\ar@{.>}[dd]\ar@{.>}[uull]\ar@{.>}[uurr]
&&&\boxed{\txt{recurrence\\solving}}\ar[dl]\\
&\txt{indefinite\\ 
integration}\ar[drr]&&&&\txt{indefinite\\summation}\ar[dll]&\\
&&&\ar@/^2.5pc/[uulll]\boxed{\txt{special\\function\\algorithms}}\ar@/^-2.5pc/[uurrr]&&&\\
}$$
\normalsize
\caption{\label{fig:tools} The computer algebra  and special function tools in 
interaction}
\end{figure}
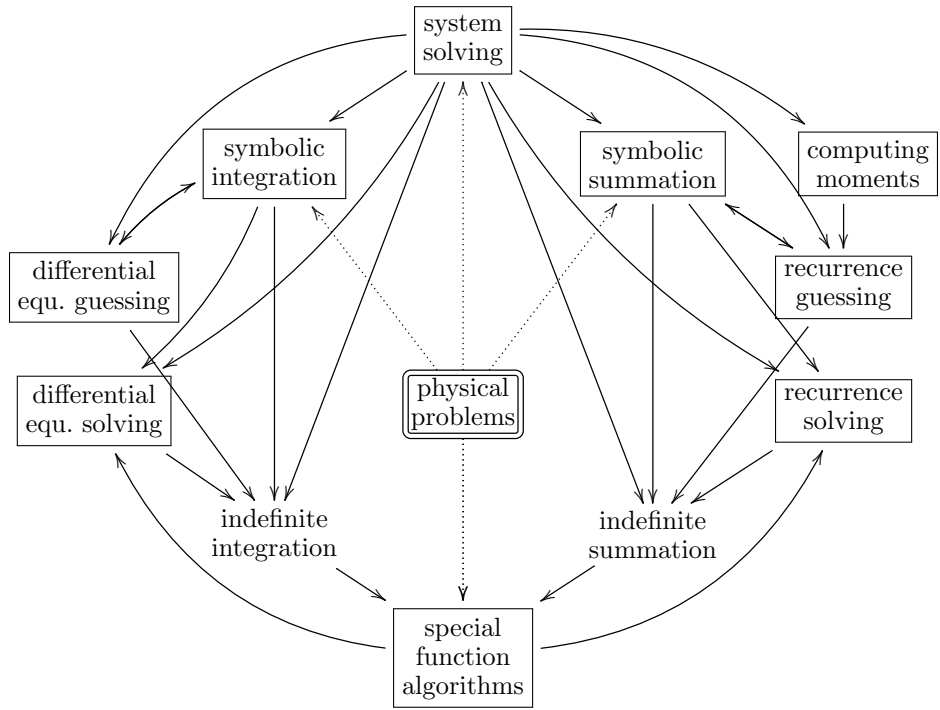

\vspace*{1mm}
\noindent
For less than the last quarter century, a technological revolution has happened in the field of perturbative 
analytic calculations in renormalizable Quantum Field Theories, which is accompanying this field since.
Considering single scale Feynman diagrams, in the time until 1998, the analytic integration of these 
amplitudes has been an art based on hypergeometric function structures and maximally Nielsen integrals 
dealing with sets of up to $O(50)$ Feynman diagrams mostly to two--loop order. Before about this time 
computer algebra has been inspired and motivated often by its own discipline or by other mathematical 
research areas, such as combinatorics, number theory, or special functions. In this survey article we 
introduced recent methods from both communities and showed how they can be combined in non-trivial ways to new 
methods that may be instrumental for current and future precision calculations in particle physics. For a 
graphical summary of the different interactions presented in this article we refer to Fig.~\ref{fig:tools}.

Elements of the revolution in Quantum Field Theories were also (quasi)shuffle 
algebras and the discovery of a hierarchy of function spaces both in Mellin $n$ 
and momentum fraction $z$ space.
For all quantities considered one may find recurrences by applying the methods of arbitrary large moments and 
guessing, which delivers closed form equations in the first place. In the moment technology presented in 
Section~\ref{Sec:largemoment} one simply deals with rational numbers ignoring completely the possible function 
spaces that may arise there. At the end of the day, one can apply the computer algebra tools introduced above and 
obtains from this data numerical representations or even symbolic representations of the final physical problem. 
In general, we feel that such new strategies will be crucial for future calculations and we are curious to see 
how these techniques can be developed further or can be outperformed with new ideas and strategies.

Systematic mathematical methods, like difference ring theory, allowed to reveal various new structures. Nowadays 
first order factorizing difference and differential equations (or systems thereof) are fully understood.
Non-first order factorizing systems, leading to $_2F_1$-solutions, complete elliptic integrals and modular forms
are understood as well and steps in the direction of equations related to Calabi--Yau manifolds are done. Yet 
these 
are rather special systems only and Feynman diagrams can in principle cause more general, yet unknown structures
also belonging to non--first order factorizing equations. They are fascinating as such and their complex analysis
is a highly interesting topic. One may intend to derive general characteristics for these quantities \cite{ABS}.
Many of the present massless and massive three-loop problems of single and double scales could be solved by the 
technologies described in the present survey and new structures challenge further innovative mathematical 
solutions and efficient computer-algebraic implementations. 
All present achievements have in various instances been achieved by sophisticated computer algebra algorithms.
Another challenge comes from the experimental 
possibilities at future colliders, operating at high luminosity, with which the theoretical results have to cope.
All methods described do not only apply to relativistic renormalizable Quantum Field Theories, but also 
to effective field theories, e.g.\ dealing with (non--linear) Einstein general relativity in post--Newtonian and 
post--Minkowskian expansions at the classical level and various applications more, e.g.\ also to solid state 
physics. Problems with more scales do still escape complete analytic solutions at present and require more
research in the future. 

The close collaboration of theoretical physicist, mathematicians and researchers in the field of computer 
algebra led both to the use of known algorithms from quite different fields in Quantum Field Theory, but have 
also triggered new mathematical and algorithmic research. The success reached has only been possible due to this 
symbiosis. This process will continue in full strength in the future.

\section*{Acknowledgments}

\vspace*{1mm}
\noindent
We would like to thank D.~Kreimer for a discussion.
This work  was supported  by the European Union's Horizon 2020 research and innovation programme under the Marie Sk\l{}odowska-Curie grant agreement No.~764850 {\it ``\href{https://sagex.org}{SAGEX}''}. We also acknowledge support from the Austrian Science Fund (FWF)
grants SFB F50 (F5009-N15) and P33530. The Feynman diagrams have been drawn using {\tt Axodraw} \cite{Vermaseren:1994je}.\\

\providecommand{\noopsort}[1]{}
\providecommand{\newblock}{}

\end{document}